\documentclass[aps,preprint,showpacs,groupedaddress,floatfix]{revtex4}

\usepackage{graphicx}

\begin{document}

\title{Quasi-particle random phase approximation
with quasi-particle-vibration coupling: application to the Gamow-Teller
response of the superfluid nucleus $^{120}$Sn}
\author{Y. F. Niu   $^{1,2}$}
\email{nyfster@gmail.com}
\author{G. Col\`{o} $^{3,1}$}
\author{E. Vigezzi $^{1}$}
\author{C. L. Bai $^{4}$}
\author{H. Sagawa $^{5,6}$}

\affiliation{$^1$ INFN, Sezione di Milano, via Celoria 16,
I-20133 Milano, Italy}
\affiliation{$^2$ ELI-NP, ¡°Horia Hulubei¡± National Institute for Physics and Nuclear Engineering, 30 Reactorului Street, RO-077125, Bucharest-Magurele, Romania}
\affiliation{$^3$ Dipartimento di Fisica, Universit\`{a} degli Studi
di Milano, via Celoria 16, I-20133 Milano, Italy}
\affiliation{$^4$ School of Physical Science and Technology,
Sichuan University, Chengdu 610065, China}
\affiliation{$^5$ RIKEN, Nishina Center, Wako 351-0198, Japan}
\affiliation{$^6$ Center for Mathematics and Physics, University of Aizu, Aizu-Wakamatsu, Fukushima 965-8580, Japan }

\date{\today}

\begin{abstract}

We propose a self-consistent quasi-particle random phase approximation
(QRPA) plus quasi-particle-vibration coupling (QPVC) model with Skyrme interactions to describe the width and the line shape of giant resonances in open-shell
nuclei, in which the effect of superfluidity should be taken into account in both the ground state and the excited states.
We apply the new model
to the Gamow-Teller resonance in
the superfluid nucleus $^{120}$Sn, including both the isoscalar spin-triplet
and the isovector spin-singlet pairing interactions. The
strength distribution in $^{120}$Sn is well reproduced and the underlying microscopic mechanisms, related
to QPVC and also to isoscalar pairing, are analyzed in detail.

\end{abstract}
 \maketitle
\date{today}

\section{Introduction}
\label{intro}

Nuclear charge-exchange transitions correspond to the transitions from an initial state in the nucleus (N, Z) to the final states in the neighbouring nuclei (N+1, Z-1) or (N-1, Z+1) \cite{Osterfeld1992, Ichimura2006}. Among the most widely knowns, one can mention the Isobaric Analog Resonance (IAR), the Gamow-Teller Resonance (GTR), and the Spin-Dipole Resonance (SDR).
These different vibrational modes, that involve  spin and isospin
degrees of freedom, provide direct and valuable information on the
isospin- and spin-isospin-dependent parts of the effective interaction
in the nuclear medium, which are otherwise poorly constrained. Nuclear charge-exchange transitions play also a crucial role in nuclear astrophysics.
GT excitations are the dominant excitation modes in weak-interaction processes such as $\beta$ decay, electron capture, and neutrino-nucleus reactions \cite{Langanke2003,Janka2007}. $\beta$-decay half-lives set the time scale of the $r$-process, and hence determine the production of heavy elements in the universe \cite{Burbidge1957,Qian2007,Niu2013}. Electron capture governs the evolution of massive stars at the end of their last hydrostatic burning phase, and influences the dynamics of core-collapse supernovae \cite{Langanke2003,Janka2007,Paar2009,Niu2011,Fantina2012}. A very accurate knowledge of spin-isospin matrix elements is also instrumental to extract the properties of the neutrinos from the measured half-life of double-$\beta$ decay \cite{Avignone2008,Vergados2012}. Therefore, nuclear charge-exchange transitions capture the interests of researchers, both experimentalists and theorists, not only in nuclear physics but also in particle physics and astrophysics.

Whereas nuclear $\beta$-decay provides directly the values of the
nuclear matrix elements of the relevant transition operator, this is
not the case when the charge-exchange states are populated by
charge-exchange reactions such as (p,n) or ($^3$He,t).
The proportionality between the reaction cross sections at the forward angles and the
GT strength has been proven to a large extent, especially  for strong
GT transitions,
  and this has paved the way
to a direct extraction of the GT matrix elements from reaction
measurements. Yet, this procedure is not entirely free from ambiguities. Moreover, no clear proportionality has been established in the case of
higher multipoles. In such a situation, it
is of paramount importance to try to improve the predictive power
of theoretical models that can provide directly the transition strengths of
the charge-exchange states of interest throughout the nuclear chart.

Two types of microscopic approaches are widely used in the
theoretical investigation of the charge-exchange excitations, i.e., the shell model and the random-phase approximation (RPA) approach which becomes
quasi-particle RPA (QRPA) for superfluid nuclei. Due to the large configuration space,
accurate shell model calculations are not feasible for heavy nuclei away from magic numbers \cite{Koonin1997,Langanke2003,Caurier2005}. The QRPA approach can be applied to all nuclei in principle except for  a few very light systems. While phenomenological QRPA has been quite popular in the past, the self-consistent QRPA approach based on Skyrme \cite{Fracasso2005,Fracasso2007,Li2008,Bai2014,Bender2002,Sarriguren2015,Yoshida2008} or relativistic \cite{Paar2004,Paar2007,Niu2013} density functionals has become increasingly accurate and successful in reproducing the observed properties of charge-exchange excitations.

At the RPA level, the nuclear collective motion is treated as a
superposition of 1 particle-1 hole (1p-1h) excitations; in the QRPA case, this becomes a superposition of two quasi-particle (2qp) excitations.
However, the energy and angular momentum of the collective motion can be
transferred to more complicated nuclear states having 2p-2h, $\ldots$, $n$p-$n$h
character (or 4qp, $\ldots$, $n$qp character in the superfluid case).
This produces the spreading width of giant resonances. In general,
the (Q)RPA approach is not able to describe the fragmentation and the
detailed line shape
of the multipole response.
The RPA plus particle-vibration coupling (RPA+PVC) is
an extension of the RPA approach which has turned out to be quite
effective, and in which
the 1p-1h configurations are coupled to collective vibrational phonons \cite{Bortignon1977,Bertsch1983,Colo1994}.
The self-consistent RPA+PVC approach for the charge-exchange excitations has been established within both the relativistic \cite{Marketin2012,Litvinova2014} and the non-relativistic framework \cite{Niu2012,Niu2014}. In both cases, it has
been possible to show that a conspicuous spreading width is developed with the inclusion of PVC effects, and thus good agreement with experimental data for the GTR and the SDR is obtained. The RPA+PVC model has also been applied to $\beta$-decay \cite{Niu2015}, and great improvement with respect to mere RPA
has been found as far as the description of the $\beta$-decay half-lives in magic nuclei is concerned.

The RPA+PVC approach is obviously limited to the case of magic nuclei.
In this paper we extend the formalism to the case of spherical superfluid nuclei,
describing the nuclear ground state within the Hartree-Fock-Bogoliubov (HFB) approximation, and the
collective excitations of the system within QRPA.
Although we will only consider the well bound nucleus $^{120}$Sn in this study, we notice that our
consistent treatment of mean-field and pairing correlations will be crucial for future
studies of exotic nuclei far from the valley of stability.
In fact, in these systems the GT strength is expected to move from the
giant resonance region to lower energies, where the transitions involving weakly bound and continuum
nucleon states play a relevant role.

We focus here mainly on the main features of the QRPA+QPVC model. As we would
like to discuss in detail the physical effects inherent in
QRPA+QPVC, we shall consider the nucleus $^{120}$Sn. This is a paradigmatic
superfluid nucleus that has been taken as a benchmark in many
calculations. In particular, it has been shown that the coupling
between quasi-particle and vibrational degrees of freedom
explains the low-lying spectra of this and of the neighbouring
nuclei in a quite convincing way (see e.g. \cite{Idini2012,Idini2015}
and references therein).

Another point is that in superfluid nuclei
both isovector (IV) and isoscalar (IS) pairing are expected to play a relevant role.
While the usual IV pairing determines the ground-state structure,
the IS pairing is present in the QRPA residual interaction for Gamow-Teller transitions. In
our previous works
\cite{Fujita2014,Bai2014,Sagawa2016}, we
have shown the importance of the GT data to pin down the value
of the IS pairing strength. Consequently, the role of such IS
pairing in calculations beyond QRPA should also be assessed.

A similar model has been recently proposed within the relativistic framework,
and applied to different giant resonances (cf. e.g. Ref.
\cite{Litvinova2008} where results for the giant dipole resonance (GDR) in
$^{120}$Sn have been presented). This model has also been applied for
the study of the GT response and the $\beta$-decay half-lives in
Ni isotopes
\cite{Litvinova2016}.
While the relativistic model is similar in
spirit to the present one, we stress again that our goal is to
address in detail the microscopic mechanisms related
to quasi-particle-vibration coupling and, to some extent, also to IS pairing.

The paper is organized as follows. In Sec. \ref{formalism} and \ref{numerical}, the formulas and
numerical details of the QRPA+QPVC model are presented. In Sec. \ref{results}, the GT
response of $^{120}$Sn
is illustrated and
a detailed analysis is provided. Finally, the main conclusions of this work
are summarized in Sec. \ref{summary}.

\section{Formalism}\label{formalism}

We first carry out a self-consistent HFB+QRPA calculation of the GT strength, using a standard Skyrme interaction. The detailed formulas
of charge-exchange QRPA on top of HFB can be found in Ref. \cite{Bai2014}. It should be noticed that
besides
the isovector $T=1$ pairing both in the ground state and in the residual interaction, the isoscalar $T=0$ pairing  must also be included in the residual interaction in
the QRPA calculation. The necessity of isoscalar $T=0$ pairing has been discussed in many previous works, especially in connection with the low-lying GT strength of $N=Z+2$ nuclei and the $\beta$-decay half-lives \cite{Bai2013,Bai2014,Fujita2014,Engel1999,Niu2013,Niu2013a,Wang2016}.
We adopt a density-dependent, zero-range surface pairing force
parameterized as follows \cite{Bai2014}:
\begin{eqnarray}
  \label{pairing}
  V_{T=1} (\mathbf{r}_1, \mathbf{r}_2) &=& V_0 \frac{1-P_\sigma}{2} (1- \frac{\rho(\mathbf{r})}{\rho_0}) \delta(\mathbf{r}_1- \mathbf{r}_2), \\
  V_{T=0} (\mathbf{r}_1, \mathbf{r}_2) &=& f V_0 \frac{1+P_\sigma}{2} (1 - \frac{\rho(\mathbf{r})}{\rho_0}) \delta(\mathbf{r}_1- \mathbf{r}_2),
\end{eqnarray}
where $\mathbf{r} = (\mathbf{r}_1- \mathbf{r}_2)/2$, $\rho_0$ is taken to be $\rho_0=0.16$ fm$^{-3}$, and $P_\sigma$ is the spin exchange operator. Although the $T=0$ pairing strength has not yet been
very firmly constrained, several different types of analysis are consistent with values of the proportionality factor $f$ which
are close to 1, or slightly larger \cite{Sagawa2016}. Accordingly, in this work we adopt the two values $f=0$ and $f=1$. This allows the reader to pin down
the effect of $T=0$ pairing, by comparing results with a typically accepted strength with results in which it has been completely neglected.

The GT excitations are
obtained by the diagonalization of the QRPA matrix.
Forward-going and backward-going amplitudes associated with the QRPA eigenstates $|n\rangle$ will be denoted by $X^{(n)}_{ab}$ and $Y^{(n)}_{ab}$, respectively.
Here and in what follows, the indices $a$, $b$ etc. label the
so-called BCS quasi-particle states in the canonical bases, that are those defined by the
operators $\alpha$ and $\alpha^\dagger$ at p. 248 of Ref. \cite{RingBook}.
Within QPVC, the QRPA strength will be shifted and redistributed through the coupling to a set of doorway states, denoted by $|N\rangle$, made of a two
BCS quasi-particle excitation $|ab\rangle$ coupled to a collective vibration $|nL\rangle$ of angular momentum $L$ and energy $\omega_{nL}$. The properties of these collective vibrations, i.e., phonons $|nL\rangle$ are, in turn, obtained by computing the QRPA response with the same Skyrme interaction, for states of natural parity $L^{\pi} = 0^+$, $1^-$, $2^+$, $3^-$, $4^+$, $5^-$, and $6^+$.
We have retained the phonons with energy less than 20 MeV and absorbing a fraction of 
the non-energy weighted isoscalar or isovector sum rule (NEWSR) strength larger than $5\%$.

The GT strength associated with QRPA+QPVC, is given by
\begin{equation}
\label{strength}
  S(E)
  = -\frac{1}{\pi} {\rm Im} \sum_\nu \langle 0 | \hat O_{\rm GT^\pm} | \nu \rangle
  ^2 \frac{1}{E - \Omega_\nu + i(\frac{\Gamma_\nu}{2}+\Delta)},
\end{equation}
where the GT operator is $\hat O_{\rm GT ^{\pm}} = \sum_{i=1}^{A} \mathbf{\sigma}(i) t_{\pm}(i)$ and $\Delta$ is a smearing parameter. In our calculation, we will only focus on the GT$^-$ excitations. $|\nu\rangle$ denote the eigenstates [associated with the complex eigenvalues
$\Omega_\nu - i\frac{\Gamma_\nu}{2}$ and eigenvectors $(F^{(\nu)}, \bar{F}^{(\nu)})$] that
are obtained by diagonalizing the energy-dependent complex matrix


\begin{equation}
  \label{PVCmatrix}
    \left( \begin{array}{cc} {\cal D} + {\cal A}_1(E) & {\cal
    A}_2(E) \\ -{\cal A}_3(E) & -{\cal D} - {\cal
    A}_4(E) \end{array} \right) \left( \begin{array}{c}
    F^{(\nu)} \\ \bar{F}^{(\nu)} \end{array} \right) = (\Omega_\nu - i
    \frac{\Gamma_\nu}{2}) \left( \begin{array}{c}
    F^{(\nu)} \\ \bar{F}^{(\nu)} \end{array} \right).
\end{equation}
${\cal D}$ is a diagonal matrix containing the physical QRPA eigenvalues.
The ${\cal A}_i$ matrices are complex and energy dependent, associated with the coupling to the doorway states.
The expressions of ${\cal A}_i$ in the QRPA basis $|n\rangle$ are given by
\begin{eqnarray}
\label{eqA}
  ({\cal A}_1) _{mn} &=& \sum_{ab,a'b'} W^\downarrow _{ab,a'b'} (E) X_{ab}^{(m)} X^{(n)}_{a'b'}
  + W^{\downarrow*} _{ab,a'b'} (-E) Y_{ab}^{(m)} Y^{(n)}_{a'b'}, \\
  ({\cal A}_2) _{mn} &=& \sum_{ab,a'b'} W^\downarrow _{ab,a'b'} (E) X_{ab}^{(m)} Y^{(n)}_{a'b'}
  + W^{\downarrow*} _{ab,a'b'} (-E) Y_{ab}^{(m)} X^{(n)}_{a'b'}, \\
  ({\cal A}_3) _{mn} &=& \sum_{ab,a'b'} W^\downarrow _{ab,a'b'} (E) Y_{ab}^{(m)} X^{(n)}_{a'b'}
  + W^{\downarrow*} _{ph,p'h'} (-E) X_{ab}^{(m)} Y^{(n)}_{a'b'}, \\
  ({\cal A}_4) _{mn} &=& \sum_{ab,a'b'} W^\downarrow _{ab,a'b'} (E) Y_{ab}^{(m)} Y^{(n)}_{a'b'}
  + W^{\downarrow*} _{ab,a'b'} (-E) X_{ab}^{(m)} X^{(n)}_{a'b'},
  \label{eqA4}
\end{eqnarray}

To speed up the calculation,
we will include in the calculation only states (in both $T_-$ and $T_+$ channels) associated with a transition strength larger than a given threshold. Note that the $T_-$ and $T_+$ channels are coupled in the QRPA and QRPA+QPVC matrices, when both protons and neutrons
are superfluid, at variance with the case of
RPA and RPA+PVC (and with the case in which only one of the two species is superfluid, as in $^{120}$Sn).
The matrix $\left( \begin{array}{cc} {\cal D} + {\cal A}_1(E) & {\cal
A}_2(E) \\ {\cal A}_3(E) & {\cal D} + {\cal
A}_4(E) \end{array} \right) $ is still symmetric as in the RPA+PVC case.

The spreading matrix $W^\downarrow _{ab,a'b'} (E)$ is the most important quantity in the QRPA+QPVC model, and it has a more general form than the
in the RPA+PVC case,
\begin{equation}
\label{Wdown}
W^{\downarrow}_{ab,a'b'} = \langle ab |V \frac{1}{E-\hat{H}} V |a'b' \rangle = \sum_{NN' }
\langle ab |V | N\rangle \langle N | \frac{1}{E-\hat{H}} | N' \rangle \langle N' | V |a'b' \rangle,
\end{equation}
 where $|N\rangle = | a'' b'' \rangle \otimes | nL \rangle$
represents a doorway state and $a'', b''$ are
BCS quasi-particle states, as recalled
above. $| nL \rangle$ is the $n-th$ phonon state with the multipolarity $L$.
The first term of Eq. (\ref{Wdown}) is
\begin{equation}
\label{eqWdown}
\langle ab | V | N \rangle = \langle ab | V | a'' b'' \otimes nL \rangle = \langle 0 | \alpha_{b} \alpha_{a} V \alpha_{a''} ^\dagger \alpha_{b''} ^\dagger \Gamma_{nL}^\dagger | 0\rangle,
\end{equation}
where $\alpha_a$ and $\alpha_a^\dagger$ are the annihilation and creation operator for the BCS quasi-particle with
quantum numbers $a \equiv \{nlj\}$,
and $\Gamma_{nL}^\dagger $ is the creation operator for phonons.
The operator $\Gamma_{nL}^\dagger $ has the following form
\begin{equation}
 \label{eqGamma}
  \Gamma_{nL}^\dagger = \frac{1}{\sqrt{1+\delta_{cd}}} \sum_{c \geq d} X^{(n)}_{cd} \alpha_{c} ^\dagger \alpha_{d} ^\dagger - Y^{(n)}_{cd} \alpha_d \alpha_c,
 \end{equation}
where $X^{(n)}_{cd}$ and $Y^{(n)}_{cd}$ are the phonon forward and backward QRPA amplitudes. Finally, we arrive at
\begin{equation}
\label{eqWdown2}
\langle ab | V | N \rangle
=  \delta_{bb''} \langle a | V | a'', nL  \rangle   +\delta_{aa''}  \langle b | V | b'', nL\rangle,
\end{equation}
where
\begin{eqnarray}
\label{eqWdown3}
  && \langle a'', nL | V | a \rangle = \langle a  | V | a'', nL \rangle \nonumber\\
  & = & \frac{1}{\sqrt{1+\delta_{cd}}} \sum_{c\geq d} \{ [   {V}_{a\bar{d}a''c}  ( u_{a''}u_cv_{ {d}}u_a -v_{ {a}''}v_{ {c}}u_dv_{ {a}}) +  {V}_{a\bar{a}''cd}  ( u_{c}u_dv_{ {a}''}u_a -v_{ {c} }v_{ {d}}u_{a''}v_{ {a}} )
    \nonumber\\
     &&+ {V}_{a\bar{c}da''}  (  u_{d}u_{a''}v_{ {c}}u_a -v_{ {d}}v_{ {a}''}u_cv_{{a}})  ]X_{cd} \nonumber \\
  && +  [   {V}_{a''\bar{d}ac}  (  u_{a }u_cv_{ {d}}u_{a''}-v_{ {a}} v_{ {c}} u_d v_{ {a}''})  +   {V}_{a''\bar{a}cd} (u_{c}u_dv_{ {a}}u_{a''}-v_{\bar{c} }v_{ {d}}u_{a}v_{ {a}''} ) \nonumber\\
  &&+  {V}_{a''\bar{c}da} ( u_{d}u_{a}v_{ {c}}u_{a''}-v_{ {d}}v_{ {a}}u_cv_{ {a}''})  ]Y_{cd}  \} .
\end{eqnarray}
The above matrix elements $V$ are calculated in the canonical basis. $v_a$ is the square root of the occupation probability for the state $a$ in the canonical basis, and $u_a = \sqrt{1-v_a^2}$ is the unoccupied amplitude. The detailed derivation of Eq. (\ref{eqWdown2}) and (\ref{eqWdown3}) can be found in Appendix A.

In the second term of Eq. (\ref{Wdown}), we have
\begin{equation}
\label{eqWsecond}
\langle N | \frac{1}{E-\hat{H}} | N' \rangle =  \langle a'' b'' \otimes nL | \frac{1}{E-\hat{H}} | a''' b'''\otimes n'L' \rangle.
\end{equation}
We will express this formula in terms of HFB quasi-particle states $|\tilde{a}\tilde{b}\rangle$  of energy
$E_{\tilde{a}} + E_{\tilde{b}} $
and we will assume
that the configurations $|\tilde{N}\rangle = |\tilde{a}\tilde{b} \otimes nL\rangle $ do not interact and are eigenstates
 of the Hamiltonian $\hat{H}$ with  eigenvalues $E_{\tilde{a}} + E_{\tilde{b}} + \omega_{nL} $.
We then obtain
\begin{equation}
\langle N | \frac{1}{E-\hat{H}} | N' \rangle = \sum_{\tilde{a}''\tilde{b}''} \frac{C_{a''\tilde{a}''}C_{b''\tilde{b}''}C^\dagger_{\tilde{a}''a'''}C^\dagger_{\tilde{b}''b'''}}{E - [E_{\tilde{a}''} + E_{\tilde{b}''} + \omega_{nL}\pm (\lambda_n-\lambda_p)] + i \Delta} ,
\end{equation}
where $C$ represents the unitary transformation matrix between HFB quasi-particle states and BCS quasi-particle states, as defined
 at p. 248 of Ref. \cite{RingBook}. The chemical potential difference $\lambda_n-\lambda_p$ is included in the energy denominator so that it can reproduce the RPA+PVC limit for magic nuclei, where the sign `$+$' is for $T_-$ excitations and and `$-$' for $T_+$
excitations. The smearing parameter $\Delta$ is introduced to avoid singularities in the denominator, and a convenient practical value is $\Delta= 200$ keV. Such a value is usually smaller than $\Gamma_{\nu}/2$ and does not affect  appreciably the QRPA+QPVC calculation of the strength in Eq. (\ref{strength}).

With the above expressions, we calculate the $W^\downarrow_{ab,a'b'}$ matrix elements, and obtain them as the sum of four terms. In
the spherical case,
we can  write all the formulas in angular momentum coupled form. The detailed derivation can be found in the Appendix B. The final expression for $W^{\downarrow J}_{ab,a'b'}$ reads
\begin{eqnarray}
\label{Wdown1}
  W^{\downarrow J}_{1ab,a'b'} &=& \delta_{j_{b} j_{b'}} \delta_{l_b l_{b'}} \delta_{j_a j_{a'}} \frac{1}{\hat{j}_a^2} \nonumber\\
   &&\sum_{a''a'''\tilde{a}''\tilde{b}''}  \sum_{nL}\delta_{j_{a''}j_{a'''}} \delta_{l_{a''}l_{a'''}}  \frac{\langle a || V || a'', nL\rangle  C_{a''\tilde{a}''}C_{b\tilde{b}''} C^\dagger_{\tilde{a}''a'''} C^\dagger_{\tilde{b}''b'} \langle a' || V|| a''',nL\rangle }{E-[E_{\tilde{a}''} + E_{\tilde{b}''} + \omega_{nL} \pm (\lambda_n-\lambda_p)] + i\Delta},
    \\
 W^{\downarrow J}_{2ab,a'b'} &=&\delta_{j_{a} j_{a'}} \delta_{l_a l_{a'}} \delta_{j_{b} j_{b'}}    \frac{1}{\hat{j}_b^2} \nonumber\\
 && \sum_{b''b'''\tilde{a}''\tilde{b}''} \sum_{nL} \delta_{j_{b''}j_{b'''}}  \delta_{l_{b''}l_{b'''}} \frac{\langle b || V || b'', nL\rangle  C_{a\tilde{a}''}C_{b''\tilde{b}''} C^\dagger_{\tilde{a}''a'} C^\dagger_{\tilde{b}''b'''} \langle b' || V|| b''',nL\rangle }{E-[E_{\tilde{a}''} + E_{\tilde{b}''} + \omega_{nL} \pm (\lambda_n-\lambda_p)] + i\Delta}, \\
  W^{\downarrow J}_{3ab,a'b'} &=&  (-)^{ j_{a}+j_{b}+J}  \left\{ \begin{array}{ccc} j_a & j_b & J \\ j_{b'} &
  j_{a'} & L \end{array} \right\}
   \sum_{a'''b''\tilde{a}''\tilde{b}''} \sum_{nL} \delta_{j_{b''}j_{b'}} \delta_{l_{b''}l_{b'}} \delta_{j_{a'''}j_{a}} \delta_{l_{a'''}l_{a}}  \nonumber\\
   && \frac{\langle b || V || b'', nL\rangle C_{a\tilde{a}''}C_{b''\tilde{b}''} C^\dagger_{\tilde{a}''a'''} C^\dagger_{\tilde{b}''b'}  \langle a' || V|| a''',nL\rangle }{E-[E_{\tilde{a}''} + E_{\tilde{b}''} +\omega_{nL} \pm (\lambda_n-\lambda_p)] + i\Delta},   \\
  W^{\downarrow J}_{4ab,a'b'} &=&  (-)^{ j_{a'}+j_{b'}+J } \left\{ \begin{array}{ccc} j_{a'} &
  j_{b'} & J \\ j_b & j_a & L \end{array} \right\}
    \sum_{b'''a''\tilde{a}''\tilde{b}''} \sum_{nL} \delta_{j_{a''}j_{a'}} \delta_{l_{a''}l_{a'}} \delta_{j_{b'''}j_{b}} \delta_{l_{b'''}l_{b}} \nonumber\\
    && \frac{\langle a || V || a'', nL\rangle C_{b\tilde{b}''}C_{a''\tilde{a}''} C^\dagger_{\tilde{b}''b'''} C^\dagger_{\tilde{a}''a'}\langle b' || V|| b''',nL\rangle  }{E-[E_{\tilde{a}''} + E_{\tilde{b}''} +\omega_{nL} \pm (\lambda_n-\lambda_p)] + i\Delta}  .
\end{eqnarray}
In the above formulas, ${\hat{j}_i^2}$ is a shorthand notation for $2j_i+1$.
The reduced matrix element has the following form
\begin{equation}
  \langle a || V || a'',nL
  \rangle = \frac{\hat{L}}{\sqrt{1+\delta_{cd}}}  \sum_{cd} [\widetilde{V}(cdLa'';a) X_{cd}^{nL} + (-1)^{j_a-j_{a''}+L} \widetilde{V}(cdLa;a'') Y_{cd}^{nL}],
\end{equation}
where
\begin{eqnarray}
  \widetilde{V}(cdLa'';a) &=& V^{L ph}_{ada''c} (u_au_{a''}u_cv_d - v_av_{a''}v_cu_d  ) + V^{Lph}_{aca''d} (u_au_{a''}v_cu_d-v_av_{a''}u_cv_d)(-)^{j_c-j_d+L} \nonumber\\
  && -V^{Lpp}_{aa''cd}(u_av_{a''}u_cu_d - v_au_{a''}v_cv_d),
\end{eqnarray}
and
\begin{eqnarray}
  \widetilde{V}(cdLa;a'') &=& V^{Lph}_{a''dac} ( u_au_{a''}u_cv_d-v_av_{a''}v_cu_d ) + V^{Lph}_{a''cad} (u_au_{a''}v_cu_d-v_av_{a''}u_cv_d)(-)^{j_c-j_d+L} \nonumber\\
  && -V^{Lpp}_{a''acd}(v_au_{a''}u_cu_d-u_av_{a''}v_cv_d).
\end{eqnarray}
This expression for $\langle a || V || a'',nL \rangle$ turns out to be in agreement with Ref. \cite{Sluys1993}.
The ph and pp interaction will take the same form as that used for non-charge-exchange QRPA calculation.
The ph and pp matrix elements $V^{L ph}_{abcd}$ and $V^{L pp}_{abcd}$ are in their angular momentum coupled form,
\begin{eqnarray}
V^{L ph}_{abcd} &=&\sum_{m_am_bm_cm_d} \langle j_a m_a j_c -m_c | L M \rangle (-)^{j_c -m_c} \langle j_d m_d j_b -m_b | L M \rangle (-)^{j_b-m_b} V^{ph}_{abcd}, \\
V^{L pp}_{abcd} &=& \sum_{m_am_bm_cm_d} \langle j_a m_a j_b m_b | L M \rangle  \langle j_c m_c j_d m_d | L M \rangle V^{pp}_{abcd}.
\end{eqnarray}

\begin{figure}[htp]
\centerline{
\includegraphics[scale=0.65,angle=0]{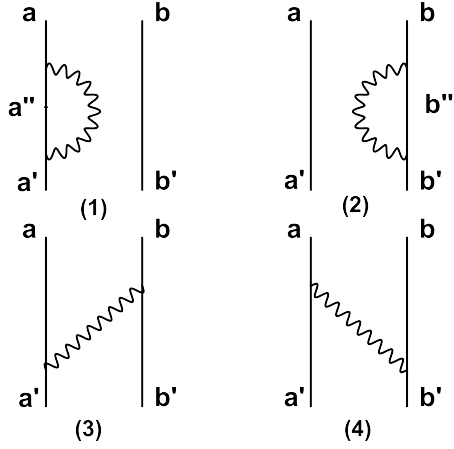}
} \caption{Diagrammatic representation of the four terms whose sum
gives the matrix element $W^\downarrow _{ab,a'b'}$.
} \label{fig1}
\end{figure}
%

For nuclei not far from the stability line, like the nucleus $^{120}$Sn
studied in this work,
the BCS quasi-particle states represent a convenient and accurate
approximation to the HFB states.
The corresponding  expression for the spreading matrix elements is obtained by approximating
the $C-$transformation with the identity, that is,
putting $C_{a\tilde a''}= \delta_{a\tilde a''}$ in Eqs. (16-19). One then obtains
\begin{eqnarray}
\label{Wdown2}
W^{\downarrow J}_{1ab,a'b'} &=&\delta_{bb'}\delta_{j_a j_{a'}} \frac{ 1  }{\hat{j}_a^2}  \sum_{a'',nL}
  \frac{\langle a || V || a'',nL
  \rangle \langle a' || V || a'',nL
  \rangle} {E-[\omega_{nL}+E_{a''}+E_b \pm (\lambda_n - \lambda_p)]+i\Delta}
  ,\nonumber\\
W^{\downarrow J}_{2ab,a'b'} &=& \delta_{ a {a'}}   \delta_{j_b j_{b'}}\frac{1  }{\hat{j}_b ^2} \sum_{b'',nL}
  \frac{\langle b || V || b'',nL
  \rangle \langle b'  || V || b'',nL
  \rangle } {E-[\omega_{nL}+E_{b''}+E_a \pm (\lambda_n - \lambda_p)]+i\Delta} ,\nonumber\\
W^{\downarrow J}_{3ab,a'b'} &=& (-)^{ j_{a}+j_{b}+J} \left\{ \begin{array}{ccc} j_a & j_b & J \\ j_{b'} &
  j_{a'} & L \end{array} \right\} \sum_{nL}
  \frac{\langle a' || V || a, nL  \rangle \langle b  || V || b',nL
   \rangle }{E-[\omega_{nL}+E_{a}+E_{b'}\pm (\lambda_n - \lambda_p)]+i\Delta}
   , \nonumber\\
W^{\downarrow J}_{4ab,a'b'}&=& (-)^{ j_{a'}+j_{b'}+J } \left\{ \begin{array}{ccc} j_{a'} &
  j_{b'} & J \\ j_b & j_a & L \end{array} \right\} \sum_{nL}
  \frac{\langle a || V || a',nL \rangle  \langle b' || V || b , nL
  \rangle}{E-[\omega_{nL}+E_{a'}+E_{b}\pm (\lambda_n - \lambda_p)]+i\Delta}
   ,
\end{eqnarray}
where $E_a$ is the BCS quasi-particle energy. The four terms correspond to the four diagrams in Fig. \ref{fig1}. These formulas are in agreement with the
formulas in Ref. \cite{Litvinova2008} and \cite{Litvinova2016}.


\section{Numerical Details}
\label{numerical}

The HFB code introduced in Ref. \cite{Bennaceur2005} is used for the calculation of ground-state properties. The HFB equations are solved in coordinate space on a radial mesh of size $0.1$ fm, within a spherical box having a radius equal to $20$ fm.
The pairing strength is determined by reproducing the neutron pairing gap in $^{120}$Sn, which is $\Delta_n =1.34$ MeV.

\begin{figure}[htb]
\includegraphics[scale=0.35]{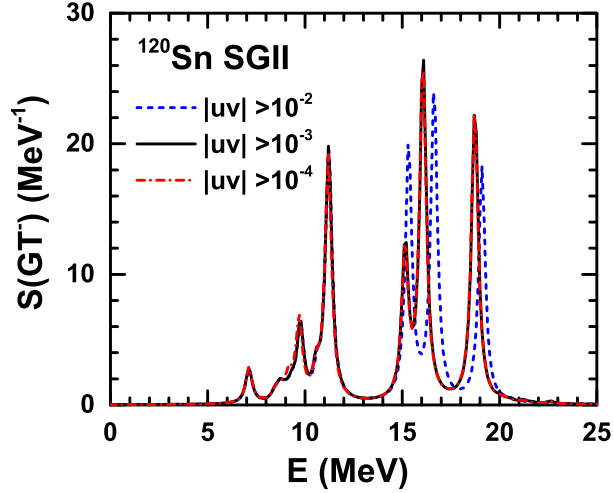}
\caption{
(Color online) GT strength function calculated with the interaction SGII \cite{Giai1981} within the QRPA approach, using different configuration spaces associated with
different threshold values for the product $|uv|$ of the occupation amplitudes in canonical basis.}
\label{fig2}
\end{figure}

\begin{figure}[htb]
\includegraphics[scale=0.3]{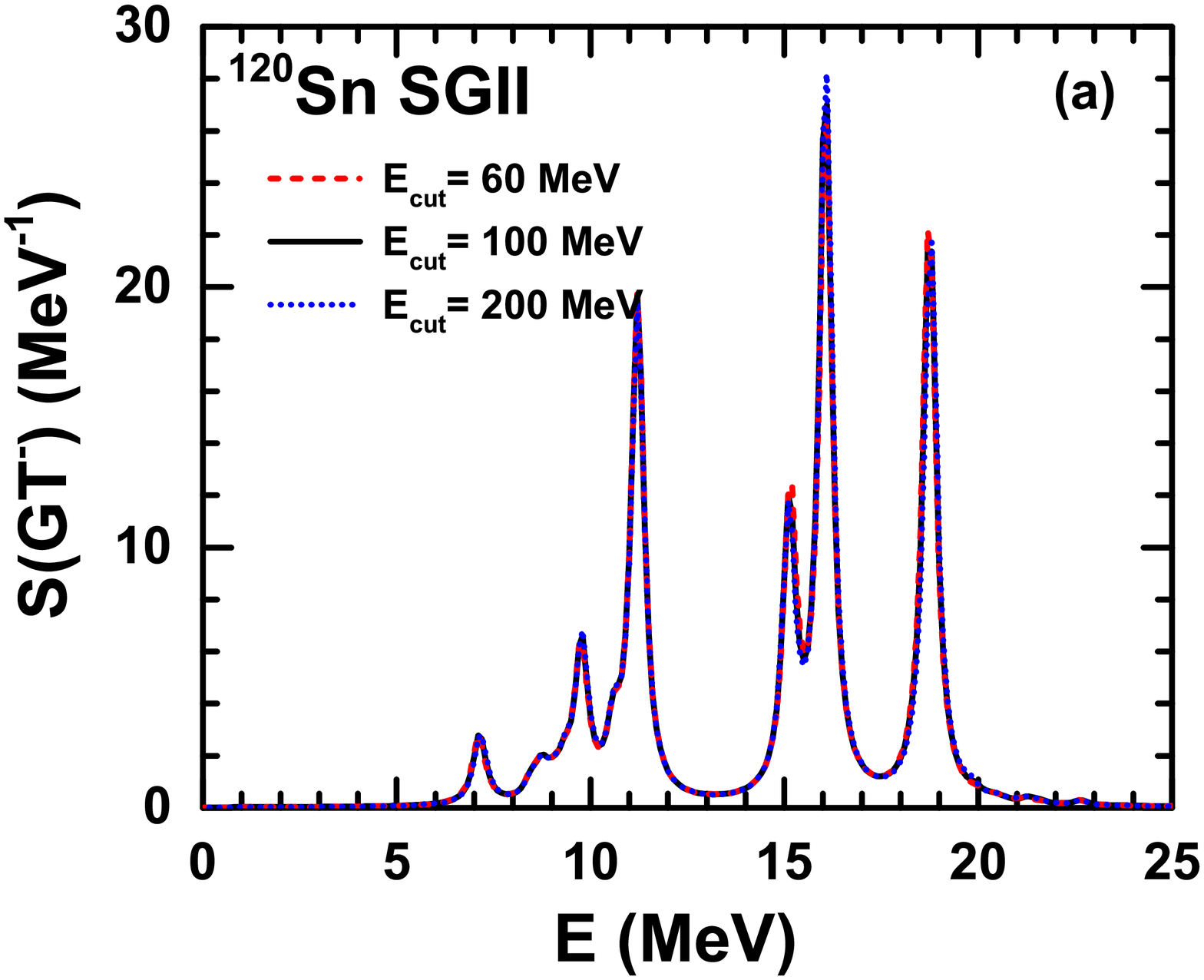}
\includegraphics[scale=0.3]{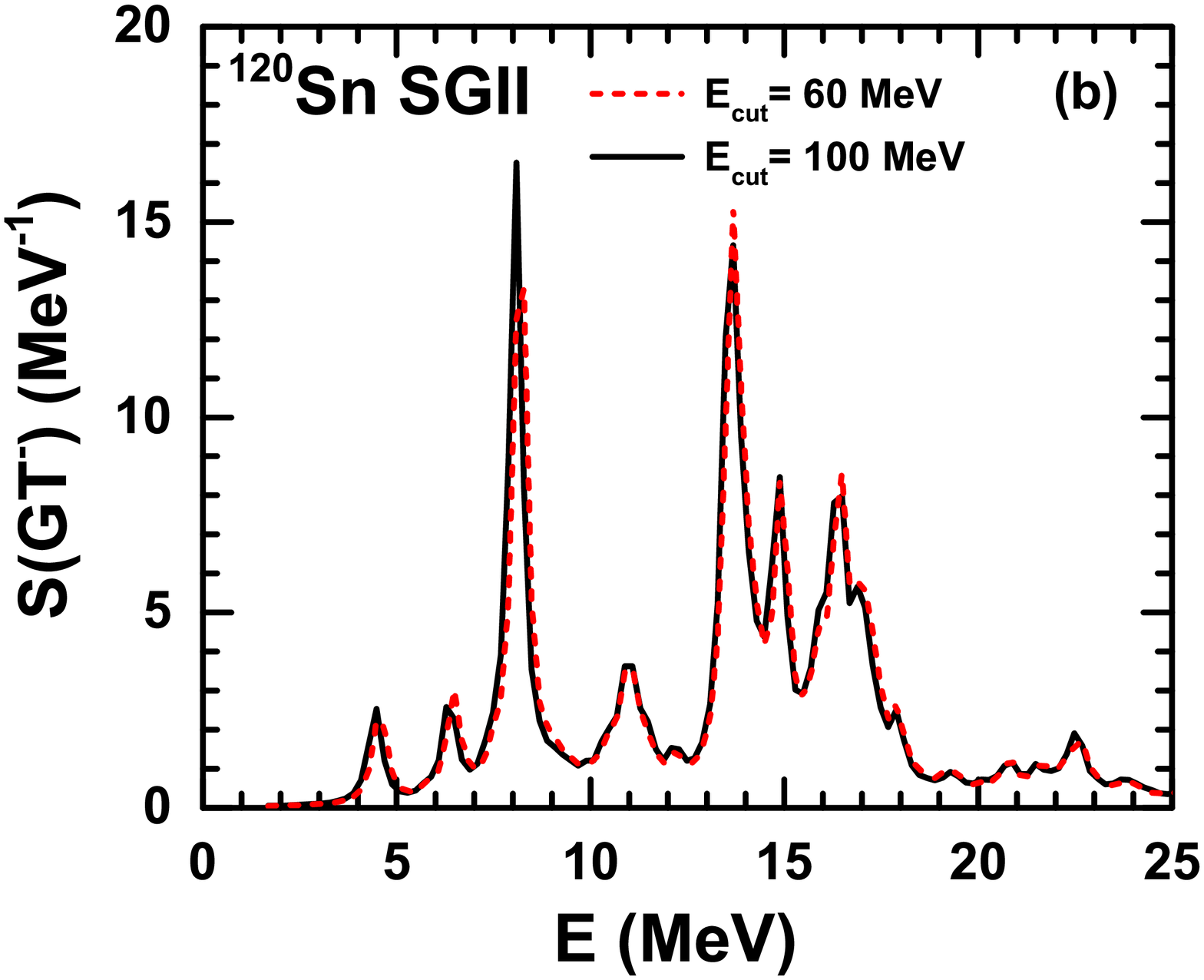}
\caption{
(Color online)
GT strength function calculated with the interaction SGII within QRPA [panel(a)] or QRPA+QPVC [panel(b)]
with different configuration spaces associated with
different values of the energy cutoff $E_{cut}$.
}
\label{fig3}
\end{figure}

The configuration space for the QRPA calculation
is defined by selecting two quasi-particle states $a$ and $b$ associated with an absolute value of the product $|u_a v_b|$ or $|u_b v_a|$ (denoted as $|uv|$) larger than a given lower cutoff,
and with quasi-particle energies smaller than $E_{cut}$. The same value of $E_{cut}$ is used for the pairing window in the HFB calculation and the intermediate states of diagrams in Fig. \ref{fig1} in the  QPVC calculation. In order to check the influence of the configuration space, by taking the GT response of $^{120}$Sn calculated with the Skyrme interaction SGII as an example, we performed a test of the convergence with respect to the parameters $|uv|$ and $E_{cut}$ (cf. Fig. \ref{fig2} and Fig. \ref{fig3}, respectively). From Fig. \ref{fig2}, we can see that when the threshold for the product $|uv|$ is smaller than $ 10^{-3}$, the GT strength distribution is quite stable.
Accordingly, the value $|uv| = 10^{-3}$ will be adopted as a lower limit in our calculations.
In Fig. \ref{fig3}, panels (a) and (b), we check the convergence
of the GT strength distribution with respect to $E_{cut}$ calculated, respectively, within QRPA and QRPA+QPVC. The results are stable for $E_{cut}$ larger than 60 MeV. The value
$E_{cut}=100$ MeV will be used in the calculations in Sec. \ref{results}.
Within the present section, to save  computation time we use $E_{cut}=60$ MeV.

\begin{figure}[htb]
\includegraphics[scale=0.35]{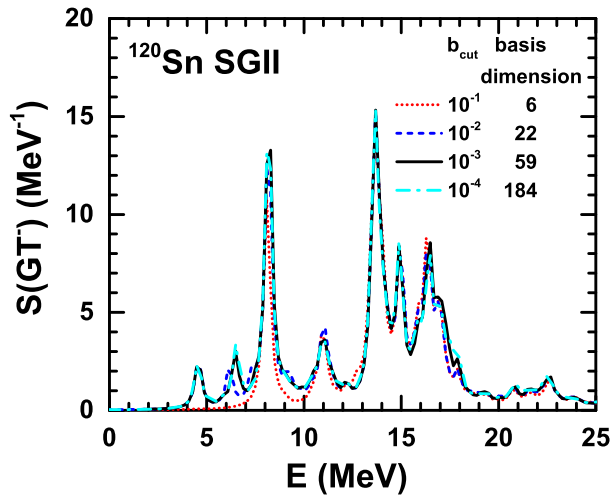}
\caption{
(Color online)
GT strength function calculated with the interaction SGII within QRPA+QPVC
with different configuration spaces associated with
different values of the strength cutoff $b_{cut}$.
}
\label{fig5}
\end{figure}

In the QPVC calculation, Eq. (\ref{PVCmatrix}) is solved in the QRPA basis. In order to simplify the calculation, we usually neglect QRPA states with very small GT strength,
reducing significantly the dimension of the QRPA+QPVC matrix. The influence of the reduction of QRPA basis on the final GT strength distribution is checked in Fig. \ref{fig5}.
The cutoff on the relative strength of the QRPA states is denoted as $b_{cut}$, namely
only the QRPA states with a fraction of NEWSR strength larger than $b_{cut}$ are included in the calculation.
We will adopt the value $b_{cut}=0.001$, which is sufficient for convergence, as shown in Fig. \ref{fig5}.

\begin{figure}[htb]
\includegraphics[scale=0.35]{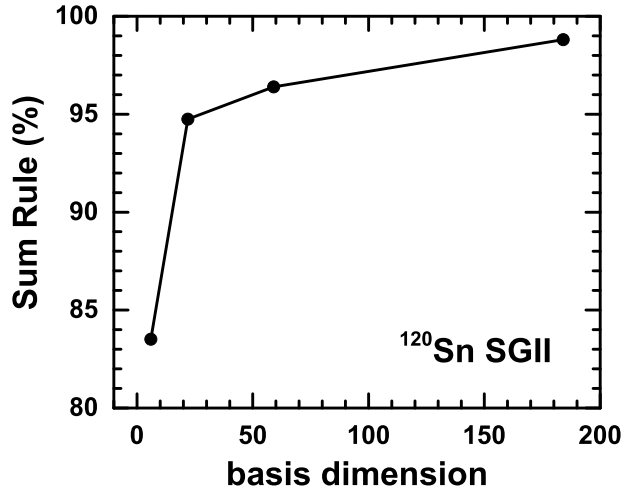}
\caption{
(Color online) Ikeda sum rule fulfillment as a function of the number of QRPA basis states used for the QPVC calculation,
in the case of the Gamow-Teller response of $^{120}$Sn calculated with the interaction SGII.
}
\label{fig4}
\end{figure}

Finally, we check the sum rule in $^{120}$Sn, within the QRPA+QPVC calculations, as a function of the number of QRPA basis states obtained
by setting $b_{cut}=10^{-1}, 10^{-2}, 10^{-3}$ and $10^{-4}$, in Fig. \ref{fig4}. We consider the integrated strength up to the excitation energy of $80$ MeV.
For $b_{cut}= 10^{-3}$, we obtain 97\% of the Ikeda sum rule.

\begin{figure}[htb]
\includegraphics[scale=0.35]{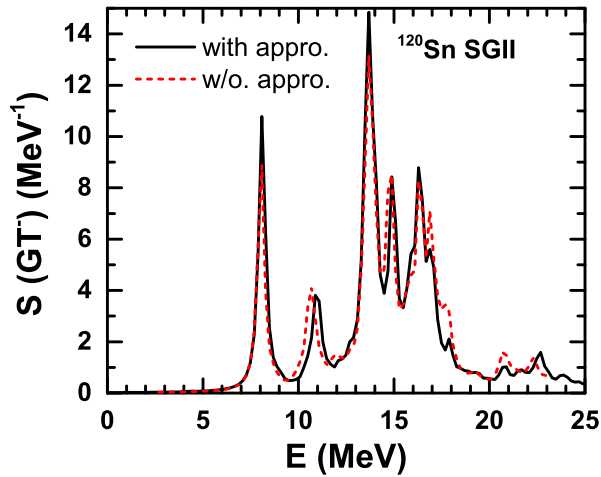}
\caption{
(Color online) Gamow-Teller strength distribution in $^{120}$Sn calculated by means of the Skyrme QRPA+QPVC model with and without the approximation of Eq.
(\ref{Wdown2})
in the spreading matrix elements.
}
\label{Fig0}
\end{figure}

Previous calculations have been made using the approximation Eq. (\ref{Wdown2}) for the spreading matrix elements. Its validity is checked in Fig. \ref{Fig0} through the comparison between the results with and without the approximation for $^{120}$Sn using the Skyrme interaction SGII. In order to save
computation time, in this case we use $b_{cut}=0.1$ instead of $b_{cut}=0.001$.
It turns out the two results are in very
good agreement with each other.
Therefore, in the following QRPA+QPVC calculations, we will use the approximation (\ref{Wdown2}).

In the calculations presented in this section, we have included the isoscalar pairing in QRPA, and we did not adopt the subtraction method
in QRPA+QPVC: this topic will be discussed in the next section. The excitation energies are always referred to the mother nucleus.

\section{Results and Discussions}
\label{results}

Before we proceed with the discussion of our results, we would like to introduce the so-called ``subtraction'' method.
The parameters of the energy density functional (EDF) are optimised so as to reproduce in the best way nuclear ground state properties, and therefore
 ``static'' correlations are implicitly taken into account. When processes beyond mean field are explicitly considered in extended RPA approaches based on this EDF,  the parameters of the EDF should be in principle readjusted to avoid problems of double counting \cite{Moghrabi2010}. This is usually not done, and as an alternative method to avoid the double counting of static correlations, it has been proposed to subtract the energy-independent part of the self-energy ~\cite{Tselyaev2007}. Recently, it has also been found that this procedure guarantees the validity of the stability condition in extensions of the RPA approach \cite{Tselyaev2013}. The theoretical foundation and application of the subtraction method
were further discussed within the formalism of second RPA in Ref. \cite{Gambacurta2015}.

In the following, we will present results obtained with the subtraction method (while we did not use it in our previous works \cite{Niu2011,Niu2013,Niu2015}).
We correspondingly modify the QRPA+QPVC equation (\ref{PVCmatrix}), by writing
\begin{equation}
  \label{PVCmatrix_sub}
    \left( \begin{array}{cc} {\cal D} + {\cal A}_1(E) -{\cal
    A}_1(0) & {\cal
    A}_2(E) - {\cal
    A}_2(0) \\ -{\cal A}_3(E) + {\cal
    A}_3(0) & -{\cal D} - {\cal
    A}_4(E) +{\cal
    A}_4(0) \end{array} \right) \left( \begin{array}{c}
    F^{(\nu)} \\ \bar{F}^{(\nu)} \end{array} \right) = (\Omega_\nu - i
    \frac{\Gamma_\nu}{2}) \left( \begin{array}{c}
    F^{(\nu)} \\ \bar{F}^{(\nu)} \end{array} \right),
\end{equation}
so that the above equation reduces to the QRPA equation when $E=0$.
In practice we just need to introduce the following replacements in Eqs. (\ref{eqA}-\ref{eqA4}):
\begin{equation}
W^\downarrow _{aba'b'} (E) \rightarrow W^\downarrow _{aba'b'} (E) - W^\downarrow _{aba'b'} (0), \quad W^\downarrow _{aba'b'} (-E)\rightarrow W^\downarrow _{aba'b'} (-E) - W^\downarrow _{aba'b'} (0).
\end{equation}

\begin{figure}[htb]
\includegraphics[scale=0.3]{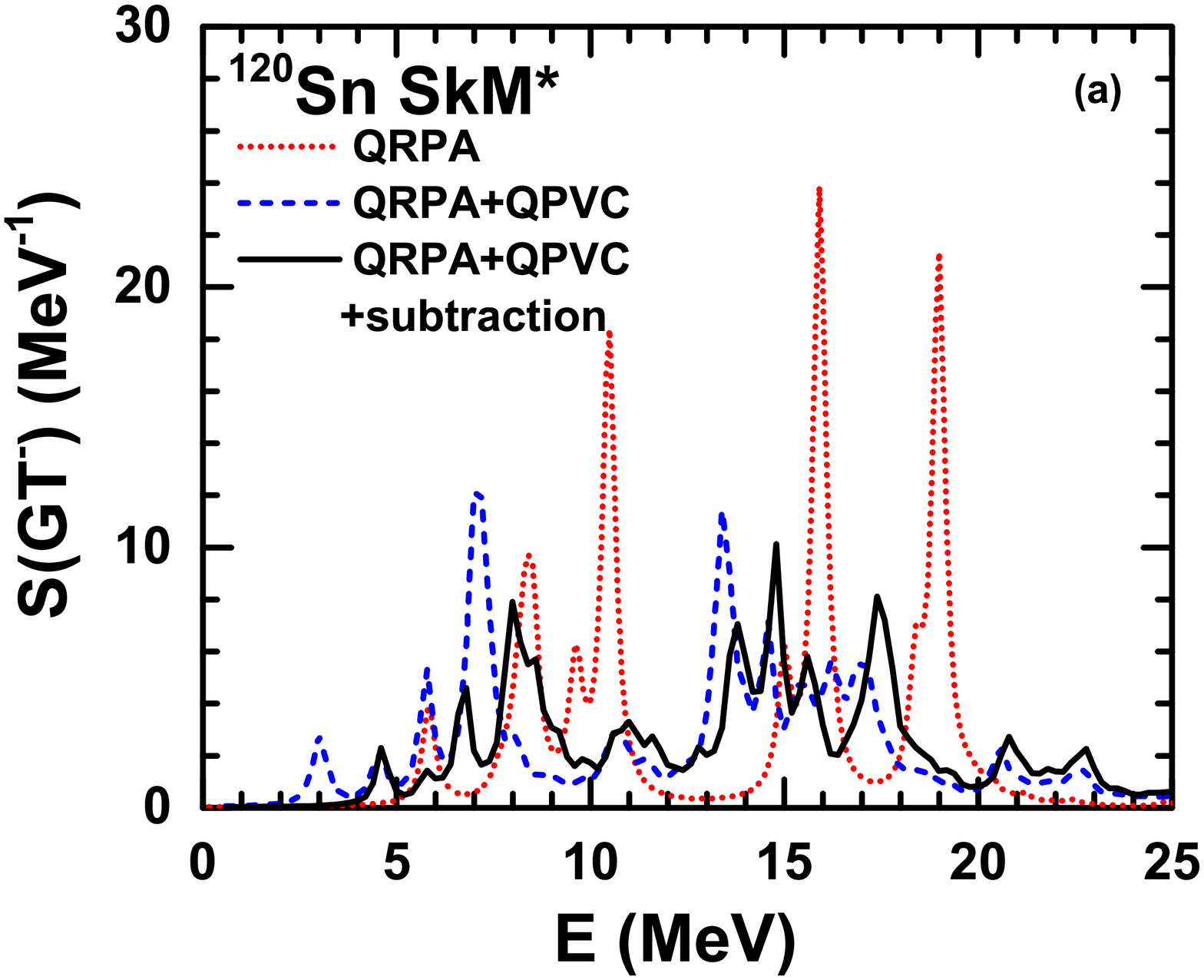}
\includegraphics[scale=0.3]{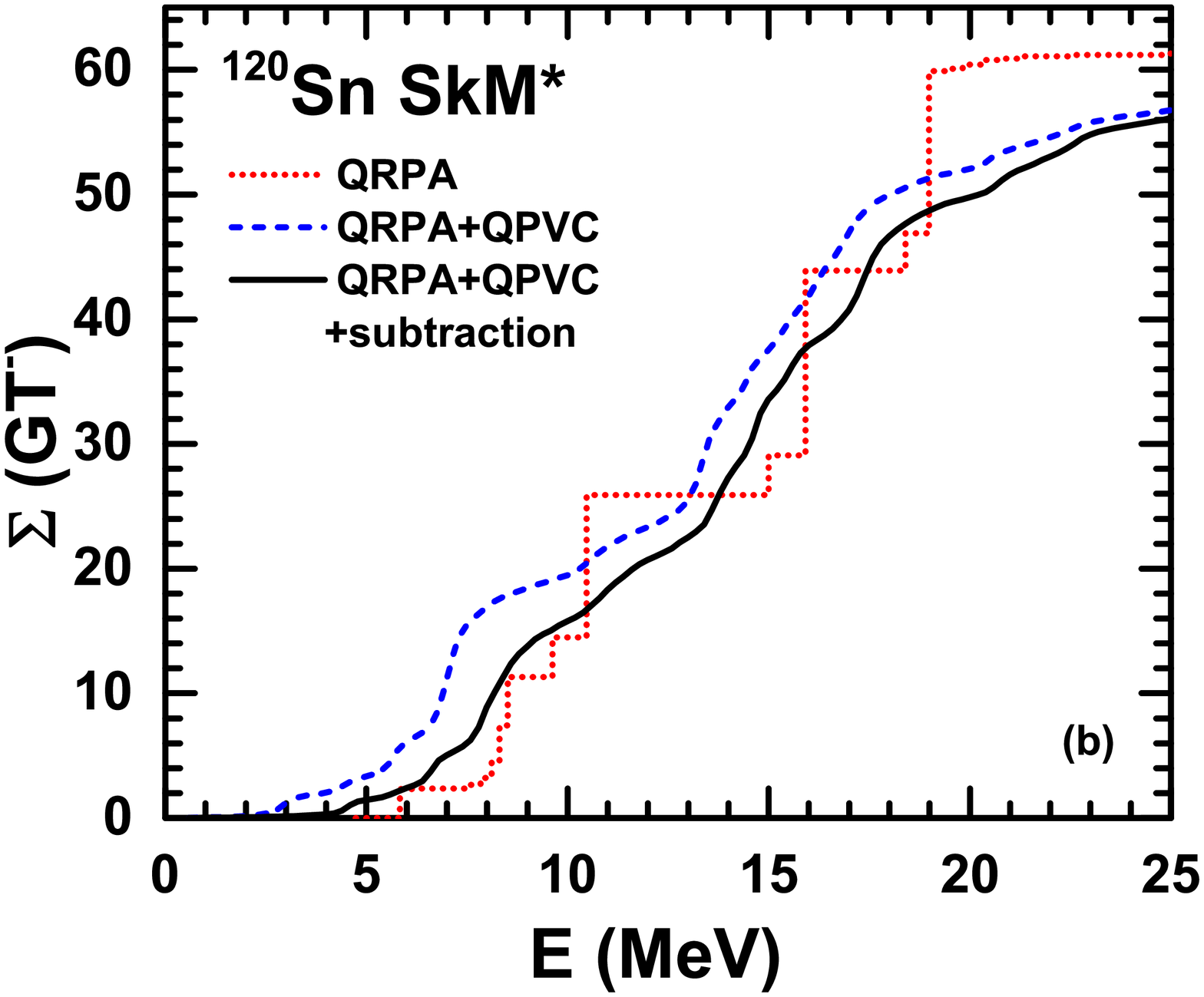}
\caption{
(Color online) The Gamow-Teller strength distributions [panel (a)] and their cumulative sums [panel (b)] for $^{120}$Sn calculated by means of QRPA and QRPA+QPVC
models, without and with subtraction method, using the Skyrme interaction SkM*.
}
\label{fig6}
\end{figure}

In Fig. \ref{fig6}, we show the effects of the subtraction method on the GT strength distribution and its cumulative sum in $^{120}$Sn, using the Skyrme interaction SkM* \cite{Bartel1982}.
Both panel (a) and panel (b) show that
 by using the subtraction method the value of real part of the self-energy is reduced, especially at low energy;
consequently, in an effective way, in introducing the subtraction method one introduces an upward shift of the
excitation energies.
The shift becomes smaller as the energy increases, and is equal to about 1 MeV in the low-energy region and to about 0.5 MeV in the giant resonance region,
until it vanishes at 25 MeV. The total GT strengths are the same for the QRPA+QPVC calculation with and without subtraction.
The width in the giant resonance region is essentially not affected,
while the width of the third low-energy peak is increased.

The values of the strength $m_0 $ and of the energy-weighted sum rule $m_1$ up to $E=25$ MeV, with and without the subtraction method, are reported in Table \ref{tablem}.
The total strengths $m_0$ obtained with and without subtraction method are very close.
The centroid energy increases by 1.2 MeV when the subtraction method is introduced, becoming very close to the
value calculated in QRPA.
This indicates that static correlations are removed by the use of the subtraction method.
In the following calculations, we will always use the subtraction method, simply indicated by the label ``QRPA+QPVC''.

\begin{table}
\caption{The strength $m_0$ and the energy weighted sum rule $m_1$, integrated up to energy $E=25$ MeV, as well as the energy centroid $m_1/m_0$ in the whole energy range $E=0-25$ MeV, calculated by QRPA, QRPA+QPVC, and QRPA+QPVC with the subtraction method in $^{120}$Sn using the interaction SkM*.}
\begin{tabular}{lccc}
    \hline\hline
        & $m_0$ & $m_1$ (MeV) & $m_1/m_0$ (MeV) \\
  \hline
  QRPA  & 61.3 & 853.4 & 13.9 \\
  QRPA+QPVC & 56.8 & 713.0 & 12.6 \\
  QRPA+QPVC+subtraction & 56.1 & 772.6 & 13.8 \\
 \hline\hline
\end{tabular}
\label{tablem}
\end{table}

\begin{table}
  \caption{The energy and reduced transition probability of the lowest phonons of different multipolarities included in the QRPA+QPVC calculation
  for $^{120}$Sn. The experimental data are taken from NNDC \cite{nndc}. The theoretical results are obtained by the QRPA approach with the interactions SAMi,
  SGII, and SkM*. }
  \begin{tabular}{ccccccccc}
    \hline\hline
            & \multicolumn{4}{c}{ $E$ (MeV) } & \multicolumn{4}{c}  { $B(EL,0\rightarrow L)$ (e$^2$ fm$^{2L}$) } \\
            \hline
    phonons& exp.    &  SAMi & SGII  & SkM*  &  exp.               &   SAMi              & SGII                & SkM*  \\
            \hline
    $2^+$   &  1.171  & 2.708 & 1.941 & 1.420 & $2.016\times 10^3$  & $1.463\times 10^3$  & $1.766\times 10^3$  & $2.632\times 10^3$ \\
    $3^-$   &
    & 3.595 & 3.313 & 3.297 &                     & $1.880 \times 10^5$ & $1.396 \times 10^5$ & $1.089 \times 10^5$ \\
    $4^+$   &
     & 4.029 & 3.757 & 3.230 &                     & $2.496\times 10^6$  & $1.568\times 10^6$  & $1.453\times 10^6$ \\
    $5^-$   &
    & 4.603 & 3.669 & 3.536 &                     & $4.454\times 10^7$  & $2.555\times 10^7$  & $3.103\times 10^7$ \\
    \hline\hline
  \end{tabular}
  \label{tablephonon}
\end{table}

We first report the properties of the collective phonons included in our QPVC calculation. The energy and reduced transition probability of the lowest phonons of different multipolarities in $^{120}$Sn, calculated by QRPA with the three Skyrme interactions SAMi \cite{Roca-Maza2012}, SGII and SkM*, are shown in Table \ref{tablephonon}. All the three interactions tend to overestimate the experimental energies, the best results being obtained with SkM*.

\begin{figure}[htb]
\includegraphics[scale=0.19]{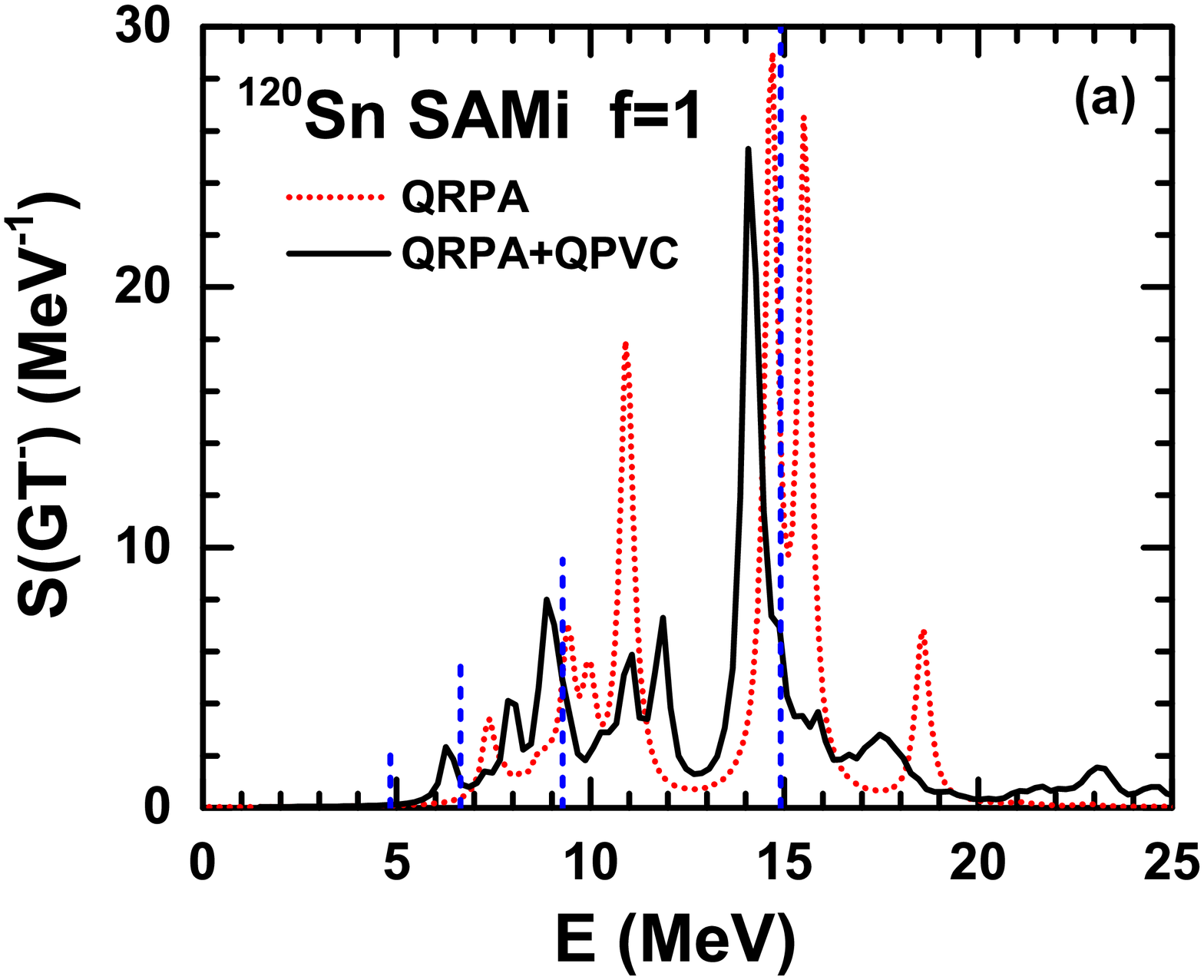}
\includegraphics[scale=0.19]{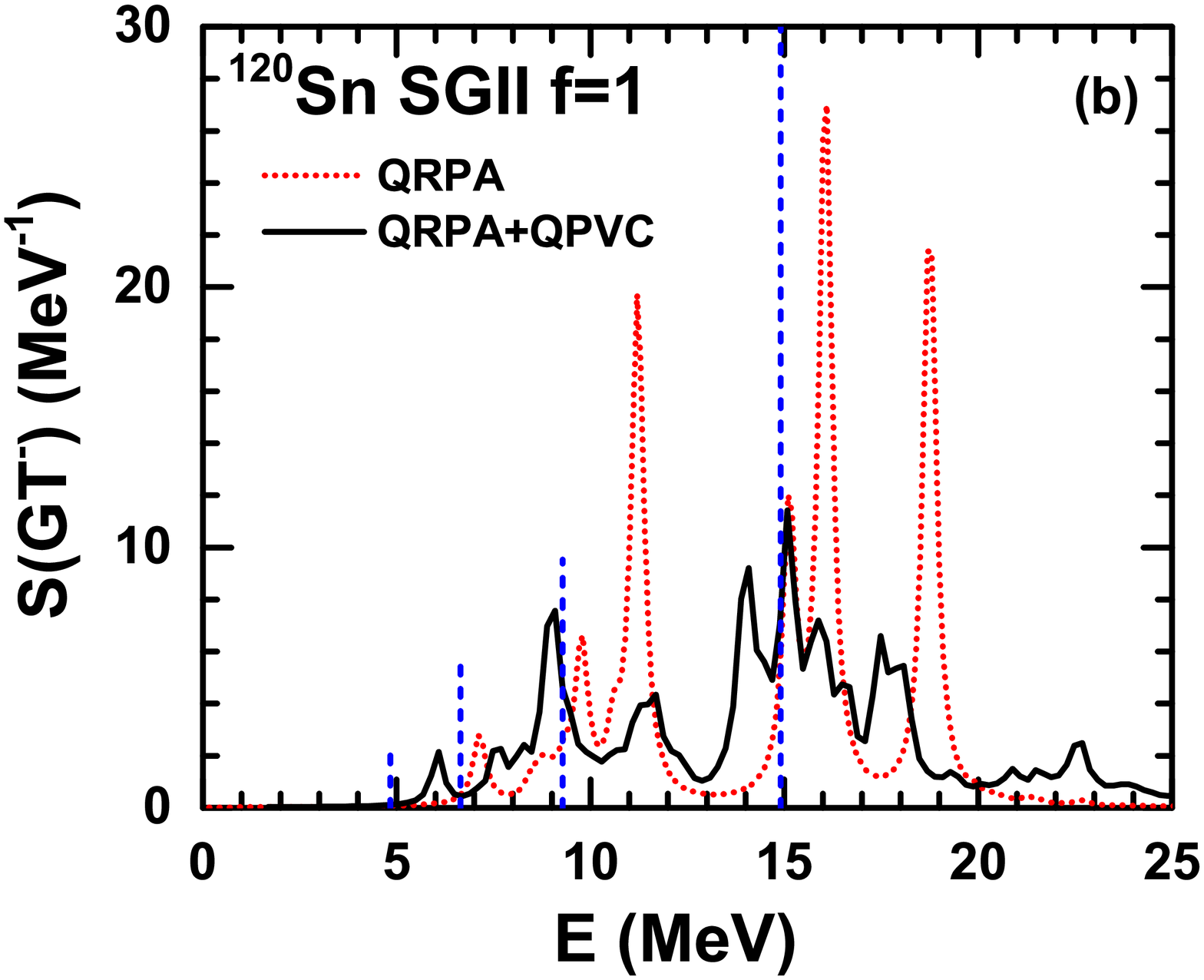}
\includegraphics[scale=0.19]{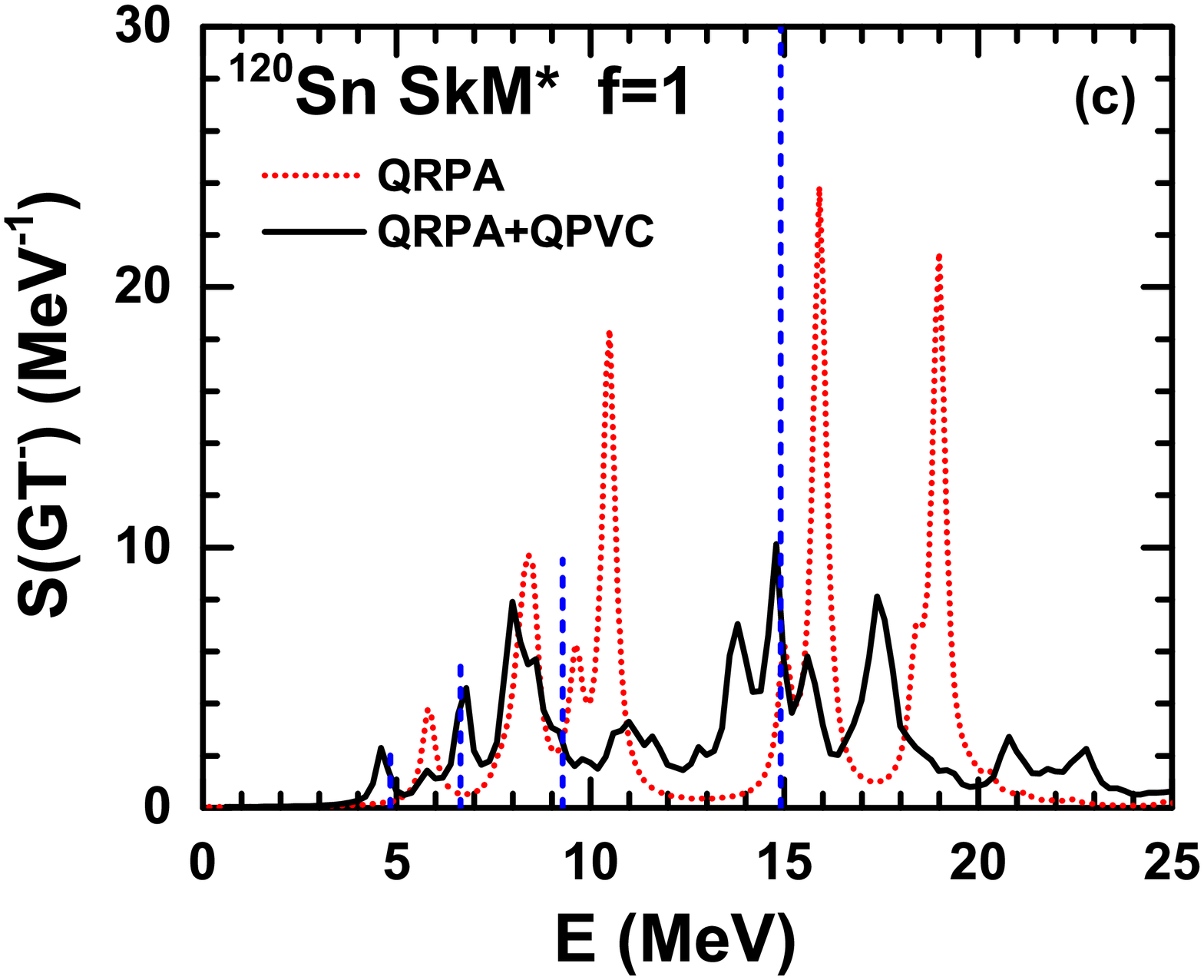}
\caption{
(Color online) The Gamow-Teller strength distributions for $^{120}$Sn calculated by QRPA and QRPA+QPVC models, with isoscalar pairing, using the interactions SAMi [panel (a)], SGII [panel (b)], and SkM* [panel (c)]. The experimental GT peak energies \cite{Pham1995} are denoted by dashed lines, and the length of each line is proportional to the cross section.
}
\label{fig7}
\end{figure}

We then show in Fig. \ref{fig7} the GT strength distributions for $^{120}$Sn calculated by the QRPA and QRPA+QPVC models
using the three Skyrme interactions,
and with a small value of the smearing parameter, $\Delta = 0.2 $ MeV.
We indicate the four peak energies identified in the ($^{3}$He, t) experiment \cite{Pham1995} by dashed lines.
The length of each line is proportional to the cross section.

With the interaction SAMi, the peaks obtained in QRPA calculation merge into a single
giant resonance peak in the QRPA+QPVC calculation. The peak is narrow, probably due to the too high phonon energies (cf. Table \ref{tablephonon}).
The GT strength distribution in the low-energy region is also redistributed and in this case some spreading width is obtained.
The QRPA calculation reproduces well the experimental giant resonance peak while the QRPA+QPVC slightly underestimates its energy.
As for the interaction SGII, the three QRPA peaks in the giant resonance region merge with the QRPA+QPVC calculation into one resonance peak with some subpeaks,
developing a spreading width of about 4.5 MeV.
We notice that if the subtraction method is not used, the width decreases to 4 MeV (cf. Fig. \ref{fig5}).
This is related to the fact that the GTR energies as well as the surface phonon energies are overestimated for this interaction at the QRPA level.
The subtraction method then improves the matching between the energy of the GTR energy and of the relevant
intermediate configurations in the calculation of the width. Although substantial, the spreading width is still smaller than the experimental value of 6.4 MeV
(cf. Figs. \ref{fig9} and \ref{fig10} below). The remaining part of the width may be due to the incorrect description of the phonon energies, to some
contribution from the escape width and to correlations coming from the coupling to other states outside our model space. In Fig. \ref{fig10}, this part of
width will be simulated by using a larger value of the smearing parameter, $\Delta=0.5$ MeV.
Besides the width, the giant resonance energy is well reproduced in the QRPA+QPVC calculation, while in the low-energy region the agreement with experimental peaks is relatively poor. The  SkM* strength distribution in the giant resonance region is quite similar to that obtained with SGII, and displays a spreading width of about 4.8 MeV.
The overall strength distribution in the low-energy region is better reproduced by SkM* than by SGII.
We will then use only the interaction SkM* in the rest of our analysis.

\begin{figure}[htb]
\includegraphics[scale=0.35]{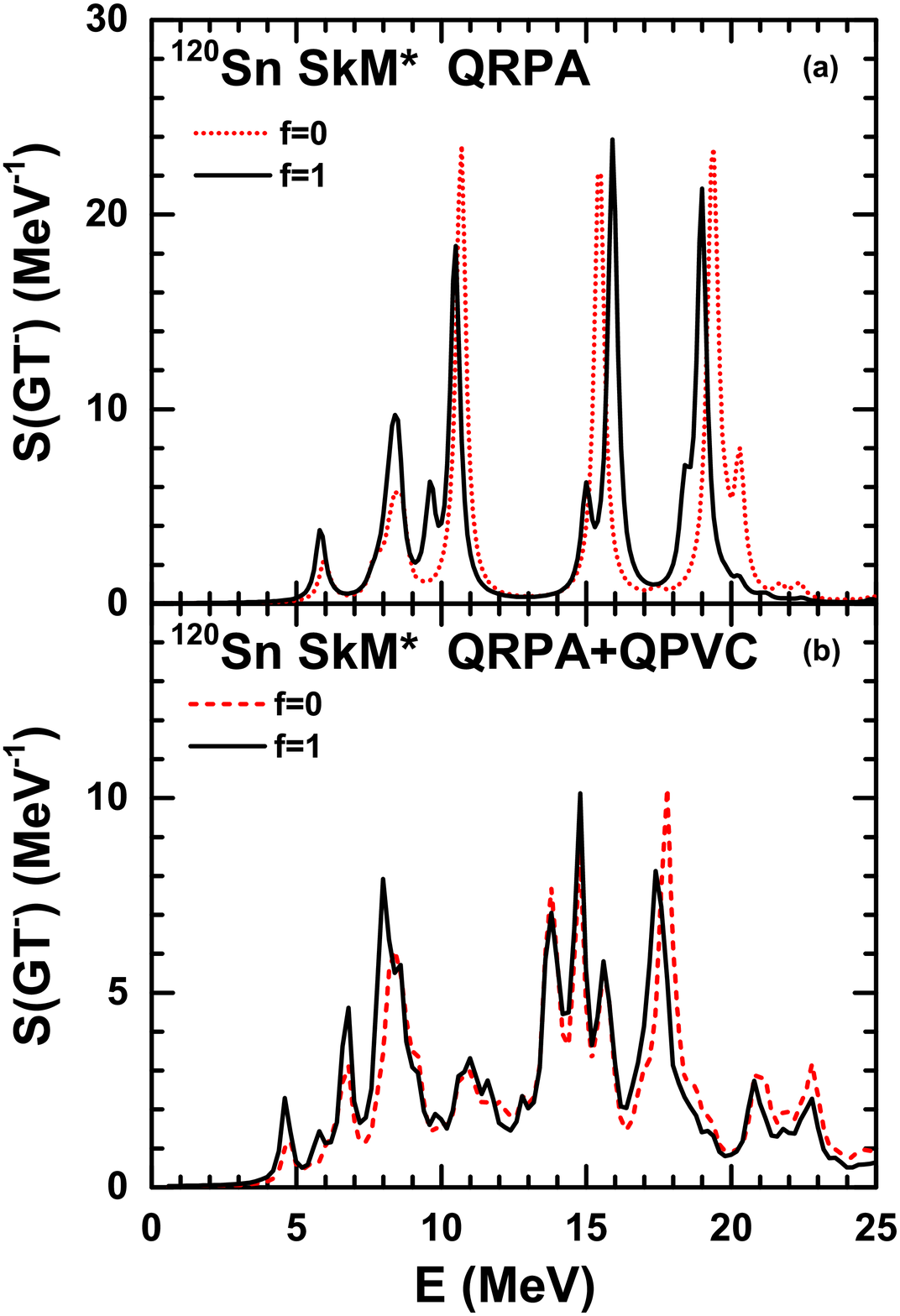}
\caption{
(Color online) The Gamow-Teller strength distributions for $^{120}$Sn calculated by QRPA [panel (a)] and QRPA+QPVC [panel (b)] models, with and without isoscalar pairing, using the Skyrme interaction SkM*.
}
\label{fig8}
\end{figure}

In Fig. \ref{fig8}, we plot the Gamow-Teller strength distributions for $^{120}$Sn calculated by the QRPA and QRPA+QPVC models, with and without isoscalar pairing. The energies and transition strength of the main GT excitations obtained in QRPA, as well as their main components, are listed in Table \ref{table2} for the case without ($f=0$) and with ($f=1$) isoscalar pairing. From Fig. \ref{fig8} as well as Table \ref{table2}, we can see that with the inclusion of the attractive isoscalar pairing the strength in the low-energy region increases; the energies of
the peaks at around 10 MeV are shifted downwards;
the splitting between the states around $E=15$ and $E=19$ MeV becomes smaller, and the strength is redistributed in favour of the lower ones.

Then we analyze in detail the microscopic structures of the main excitations in QRPA. The lowest main state lying at $E=5.96$ (5.83) MeV for $f=0$ ($f=1$) is basically a single-particle excitation of back spin-flip type, i.e., $j=l-1/2 \rightarrow j=l+1/2$.
The second lowest main state at $E=8.62$ (8.51) MeV for $f=0$ ($f=1$) is composed of several quasi-particle transitions of  non-spin-flip, i.e., $j=l\pm 1/2 \rightarrow j=l\pm 1/2$
 as well as $\Delta L=2$ type.
We note that the transitions with $\Delta L=2$ do not contribute to the GT strength, since the GT operator is characterized by $\Delta L=0$.
In the  $f=0$ case, the main components of the
state at $E=10.67$ MeV are a non spin-flip and a spin-flip transition. When $f=1$, one instead finds two states, one lying
at $E=9.62$ MeV with a strong non-spin-flip component and another at $10.47$ MeV with
a strong spin-flip component. In the giant resonance region, in the $f=0$ case, the state at $E=15.45$ MeV is composed of the two spin-flip quasi-particle transitions $\nu 1g_{9/2} \rightarrow \pi 1g_{7/2}$ and $\nu 1h_{11/2} \rightarrow \pi 1h_{9/2}$. After including the isoscalar pairing, this state splits into two states at $E=15.00$ and 15.91 MeV, and other transitions of back spin-flip are mixed into these two states. For $f=0$, the main states in the high energy region lie at $E=19.32$, $19.43$, and $20.29$ MeV, and are composed of spin-flip and non-spin-flip quasi-particle transitions. For $f=1$, the highest states lie at $E=18.40$ and $18.98$ MeV, and the strength is concentrated in the second state. The components of these states are also changed,
and the back spin-flip transition $\nu 1h_{9/2} \rightarrow \pi 1h_{11/2}$ plays an important role.

\begin{table}
  \caption{QRPA amplitudes $X^2_{ab} - Y^2_{ab}$ for GT states with large B(GT) ($>$1.0). The excitation energy $E$ and transition strength $B$ of different GT excitations in $^{120}$Sn are calculated in
  QRPA
either
   without isoscalar pairing ($f=0$) or with isocalar pairing ($f=1$). The excitation energies are given with respect to the mother nucleus in units of MeV. For each QRPA state, only the configurations
associated with the large QRPA amplitudes ($\vert X^2_{ab}- Y^2_{ab} \vert > 0.03$) are listed.}
\label{table2}
{\scriptsize
\begin{tabular}{c|ccccccc|cccccccc}
  \hline\hline
     & \multicolumn{7}{c|}{ $f=0$ }   & \multicolumn{8}{c}{  $f=1$ }  \\
     \hline
   $E$ (MeV) & $5.96$ & $8.62$ & $10.67$ & $15.45$ & $19.32$  & $19.43$ & $20.29$  &$5.83$ &$8.51$ & $9.62$ & $10.47$  & $15.00$ & $15.91$  & $18.40$ & $18.98$  \\ \hline
 $B$(GT)            &  $1.33$ & $2.47$ & $14.65$ & $14.58$ & $9.82$ & $5.87$ & $4.09$ & $2.31$ & $3.84$ & $3.14$ & $11.48$ & $3.16$ & $14.80$  & $2.96$  & $13.07$  \\ \hline
                configuration &\multicolumn{7}{c|}{ QRPA amplitude  }  &  \multicolumn{8}{c}{ QRPA amplitude  }\\
   \hline  \hline
   $\nu 2d_{3/2} \rightarrow \pi 2d_{5/2}$   &  0.94 &&& &&& & 0.97 &  &&&& &&\\
   $\nu 2d_{5/2} \rightarrow \pi 2d_{5/2}$   & &0.04 &0.04 &&&& &  &0.08 &&&& &&\\
   $\nu 2d_{5/2} \rightarrow \pi 1g_{7/2}$   & &0.15 & &&&& &  &0.11 & 0.29 &&&& &\\
   $\nu 3s_{1/2} \rightarrow \pi 3s_{1/2}$  & &0.06 & &&&& &  &0.06 &&&& &&\\
   $\nu 3s_{1/2} \rightarrow \pi 2d_{3/2}$   & &0.23 & &&&& &  &0.39 &&&& &&\\
   $\nu 2d_{3/2} \rightarrow \pi 2d_{3/2}$   & &0.34 & &&&& &  &0.23 &&&& &&\\
   $\nu 1g_{7/2} \rightarrow \pi 1g_{7/2}$   & &0.11 & &&&& &  &0.09 &&&& &&\\
   $\nu 1h_{11/2} \rightarrow \pi 1h_{11/2}$ & & &0.28 &&&& & &   & 0.61 &&&& &\\
   $\nu 2d_{5/2} \rightarrow \pi 2d_{3/2}$   & & &0.55 &&&& &  & & & 0.87 &&&& \\
   $\nu 1g_{9/2} \rightarrow \pi 1g_{7/2}$   & & &    & 0.83 &&& 0.06 &  & & &      & 0.34 &0.48& &0.08\\
   $\nu 1h_{11/2} \rightarrow \pi 1h_{9/2}$  & & &    & 0.10 & 0.59 & 0.20 & 0.06 &  & & &     & 0.18 &&0.05 &0.61\\
   $\nu 1h_{9/2} \rightarrow \pi 1h_{11/2}$  &&&&&&& &  & & &     & 0.26 &0.40 & 0.15 &0.06 \\
   $\nu 2g_{9/2} \rightarrow \pi 1g_{7/2}$   & & &    &      & 0.05 && 0.72&  & &  &    & 0.04 & &&\\
   $\nu 2f_{7/2} \rightarrow \pi 2f_{7/2}$   & & &    &      & 0.26 & 0.69 & &  & &   &   &  && 0.45 &0.09\\
   $\nu 3s_{1/2} \rightarrow \pi 5s_{1/2}$   & & &    &      &      && 0.04  &&&&&&\\
   $\nu 3d_{5/2} \rightarrow \pi 2d_{5/2}$   &&&&&&& & & &   &   &  & & 0.05 & \\
  \hline\hline
\end{tabular}
}
\end{table}

\begin{table}

  \caption{
 Microscopic structure of the main GT peaks found above $E=13$ MeV in the QRPA+QPVC calculations with IS pairing ($f=1.0$), shown in Fig. 9 (b).
 We list the peak energy $E$,
  the complex eigenenergy $\Omega_\nu - i\frac{\Gamma_\nu}{2}$ from QRPA+QPVC,
  the energy $E_m$ and the forward amplitudes  $X_{ab}^{(m)}$ of the associated QRPA state $|m\rangle$ (cf. Table  \ref{table2}),
 the imaginary part of the diagonal spreading matrix element $W_{ab,ab}$,
  and the contributions to the imaginary part of the self-energy ${\cal A}_1$.
\label{table4}}
{\scriptsize
\begin{tabular}{cc|ccc|cc}
  \hline\hline
 \multicolumn{2} {c|} {QRPA+QPVC} &\multicolumn{3} {c|} {QRPA} & \multicolumn{2} {c} {QRPA+QPVC} \\ \hline
   $E$ (MeV)& $\Omega_\nu - i\frac{\Gamma_\nu}{2}$ (MeV) &   $E_m$ (MeV) & configuration (ab) & $X^{(m)}_{ab}$ & Im $W_{ab,ab} (E)$ (MeV) & Im $({\cal A}_1)_{mm} (E)$ (MeV)  \\
   \hline
   13.79 & $(13.65 - i 0.34)$ &15.00 & $\nu 1g_{9/2} \rightarrow \pi 1g_{7/2}$ & -0.59 & -0.84 & -0.29 \\
         &    &   & $\nu 1h_{11/2} \rightarrow \pi 1h_{9/2}$ & -0.42 & -0.64 & -0.11 \\
         &    &   & $\nu 1h_{9/2} \rightarrow \pi 1h_{11/2}$ & 0.51  & -0.20 & -0.053\\
         &    &   & total & &       & -0.57 \\
         &    &15.91 & $\nu 1g_{9/2} \rightarrow \pi 1g_{7/2}$ & 0.69  & -0.84 & -0.41 \\
         &    &   & $\nu 1h_{9/2} \rightarrow \pi 1h_{11/2}$ & 0.63  & -0.20 & -0.082\\
         &    &   & total & &       & -0.49 \\
   \hline
   14.79 &  $(14.71-i0.61)$ & 15.00 & $\nu 1g_{9/2} \rightarrow \pi 1g_{7/2}$ & -0.59 & -0.64 & -0.22 \\
         &    &   & $\nu 1h_{11/2} \rightarrow \pi 1h_{9/2}$ & -0.42 & -0.52 & -0.092\\
         &    &   & total & &       & -0.58 \\
         &    & 15.91 & $\nu 1g_{9/2} \rightarrow \pi 1g_{7/2}$ & -0.69 & -0.64 & -0.31 \\
         &    &   & total & &       & -0.45 \\
   \hline
   15.59 & $(15.13-i 1.22)$    & 15.00 & $\nu 1g_{9/2} \rightarrow \pi 1g_{7/2}$ & -0.59 & -1.50 & -0.51 \\
         &    &   & $\nu 1h_{11/2} \rightarrow \pi 1h_{9/2}$ & -0.42 & -0.83 & -0.15 \\
         &    &   & total & &       & -0.95 \\
         &    & 15.91 & $\nu 1g_{9/2} \rightarrow \pi 1g_{7/2}$ & 0.69  & -1.50 & -0.72 \\
         &    &   & $\nu 1h_{9/2} \rightarrow \pi 1h_{11/2}$ & 0.63  & -0.19 & -0.076\\
         &    &   & total & &       & -0.84 \\
   \hline
   17.39 &  $(16.89 - i 0.68)$  & 15.91 & $\nu 1g_{9/2} \rightarrow \pi 1g_{7/2}$ & 0.69  & -0.88 & -0.42 \\
         &  $(16.90,-i1.14)$  &   & $\nu 1h_{9/2} \rightarrow \pi 1h_{11/2}$ & 0.63  & -0.29 & -0.11 \\
         &    &   & total & &       & -0.67 \\
         &    &18.40 & $\nu 2f_{7/2} \rightarrow \pi 2f_{7/2}$ & -0.67 & -2.59 & -1.17 \\
         &    &   & $\nu 3d_{5/2} \rightarrow \pi 2d_{5/2}$ & 0.23  & -1.38 & -0.072 \\
         &    &   & total & &       & -1.80\\
         &    & 18.98 & $\nu 1g_{9/2} \rightarrow \pi 1g_{7/2}$ & 0.28  & -0.88 & -0.067 \\
         &    &   & $\nu 1h_{11/2} \rightarrow \pi 1h_{9/2}$ & -0.78 & -0.48 & -0.30 \\
         &    &   & $\nu 2f_{7/2} \rightarrow \pi 2f_{7/2}$ & -0.30 & -2.59 & -0.23 \\
         &    &   & $\nu 2f_{7/2} \rightarrow \pi 1h_{9/2}$ & -0.15 & -2.67 & -0.063 \\
         &    &   & total & &       & -0.78 \\
\hline\hline
\end{tabular}
}
\end{table}

For the QRPA+QPVC results, the profile of the strength function in the giant resonance
region is similar in the $f = 0$ and $f = 1$ cases, although the strength of the peaks in the
low-energy region are increased and the strength of the highest peak is decreased with the
inclusion of isoscalar pairing.
In the following, we shall discuss the microscopic structure of the GTR peaks for $f=1$.

From the previous Section, we recall that
at each excitation energy $E$ we solve the QRPA+QPVC equation obtaining a set of eigenstates with complex eigenvalues $(\Omega_\nu - i\Gamma_\nu/2)$.
We focus on values $E$ corresponding to peaks in the strength function. The contribution to the width is essentially given by twice the
imaginary part of the important eigenstates that lie close to $E$.
Each of these eigenstates is a linear combination of the QRPA states $|m\rangle$ (with energy $E_m$). In the following, we will analyze the important eigenstates
and their important QRPA components. The eigenstates which give the largest contributions to the strength are given in Table \ref{table4},
together
with the main associated QRPA components $|m\rangle$. For each $|m\rangle$, the most important quasi-particle configurations
$ab$ are also listed, together with their contribution to the imaginary part of the self-energy ${\cal A}_1$. We note that the total width
$\Gamma_{\nu}$ resulting from the complete diagonalization
is different from the sum of the values of Im $({\cal A}_1)_{mm}$, due to the strong mixing between different QRPA states.

The eigenstate with the eigenvalue $(13.65 - i 0.34)$ MeV gives the most important contribution to the peak found at $E=13.79$ MeV.
This eigenstate is mainly composed of 
the QRPA states at $E=15.00$ and $15.91$ MeV (cf. Table \ref{table2}).
The contributions to the width from the imaginary parts of the self-energy of these two QRPA states are -0.57 and -0.49 MeV, respectively.
The diagrams $W_{abab}$ with $(a,b) =(\pi 1g_{7/2}, \nu 1g_{9/2})$ or $(\pi 1h_{9/2}, \nu 1h_{11/2})$ contribute most to the self-energy
and, in turn, the coupling to $2^+$ and $3^-$ phonons plays the most important role.
The same QRPA configurations give the largest contributions to the peak
with energy $E=14.79$ MeV.
At the peak energy $E=15.59$ MeV, the eigenstate with $(15.13-i 1.22)$ MeV is important, and its main components 
are again the QRPA states at $E=15.00$ and $15.91$ MeV. At this peak energy, the imaginary parts of their self-energies are increased to -0.95 and -0.84 MeV, compared to
the corresponding values -0.58 and -0.45 MeV found at the peak energy $E=14.79$ MeV. The important diagrams are still $W_{abab}$ with $(a,b) = (\pi 1g_{7/2},\nu 1g_{9/2})$ or $ (\pi 1h_{9/2},\nu 1h_{11/2}) $, but with coupling to $2^+$, $3^-$ and $4^+$ phonons.
At the peak energy $E=17.39$ MeV, the eigenstates with the eigenvalue $(16.89 - i 0.68)$ and $(16.90-i1.14)$ MeV contribute to the peak. These two states are mainly composed of the QRPA states 
 at $E=15.91$, $18.40$ and $18.98$ MeV.
The diagram $W_{abab}$ with $(a,b) = (\pi 1g_{7/2}, \nu 1g_{9/2})$, $(\pi 1h_{9/2}, \nu 1h_{11/2})  $, $ (\pi 2f_{7/2},\nu 2f_{7/2})$, or $(\pi 1h_{9/2},\nu 2f_{7/2} )$ contributes to the self-energy most, and the couplings to $2^+, 3^-, 4^+,$ and $5^-$ phonons all play important roles.

In summary, these four subpeaks are mainly composed of several QRPA states in the GTR region. For the first three peaks, the QRPA states at $E=15.00$ and $15.91$ MeV are relatively more important, while the QRPA states at $E=15.91$, $18.40$ and $18.98$ MeV, as well as the states at even higher energies with small B(GT) values, are   important for the fourth peak.

\begin{figure}[htb]
\includegraphics[scale=0.3]{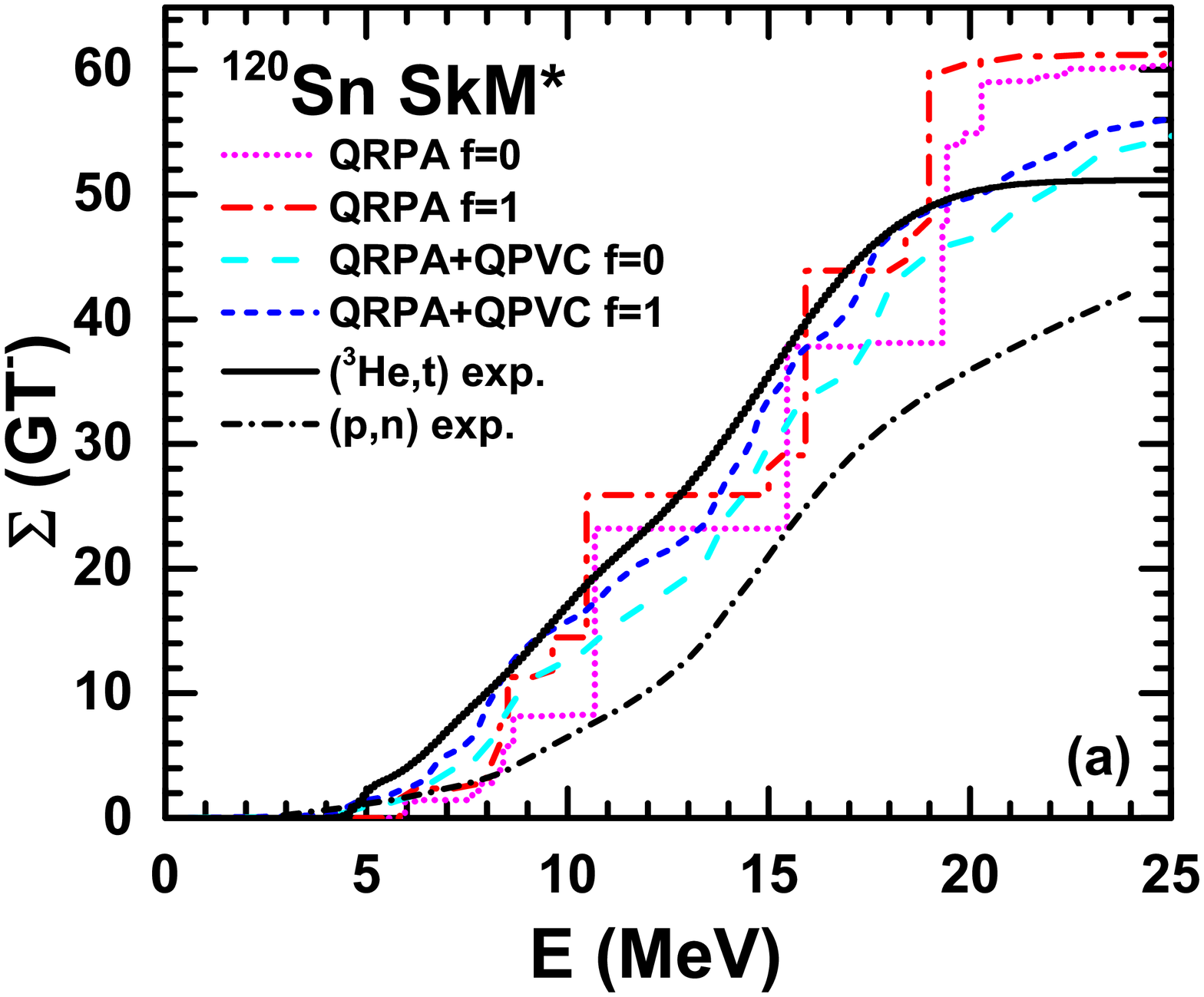}
\includegraphics[scale=0.3]{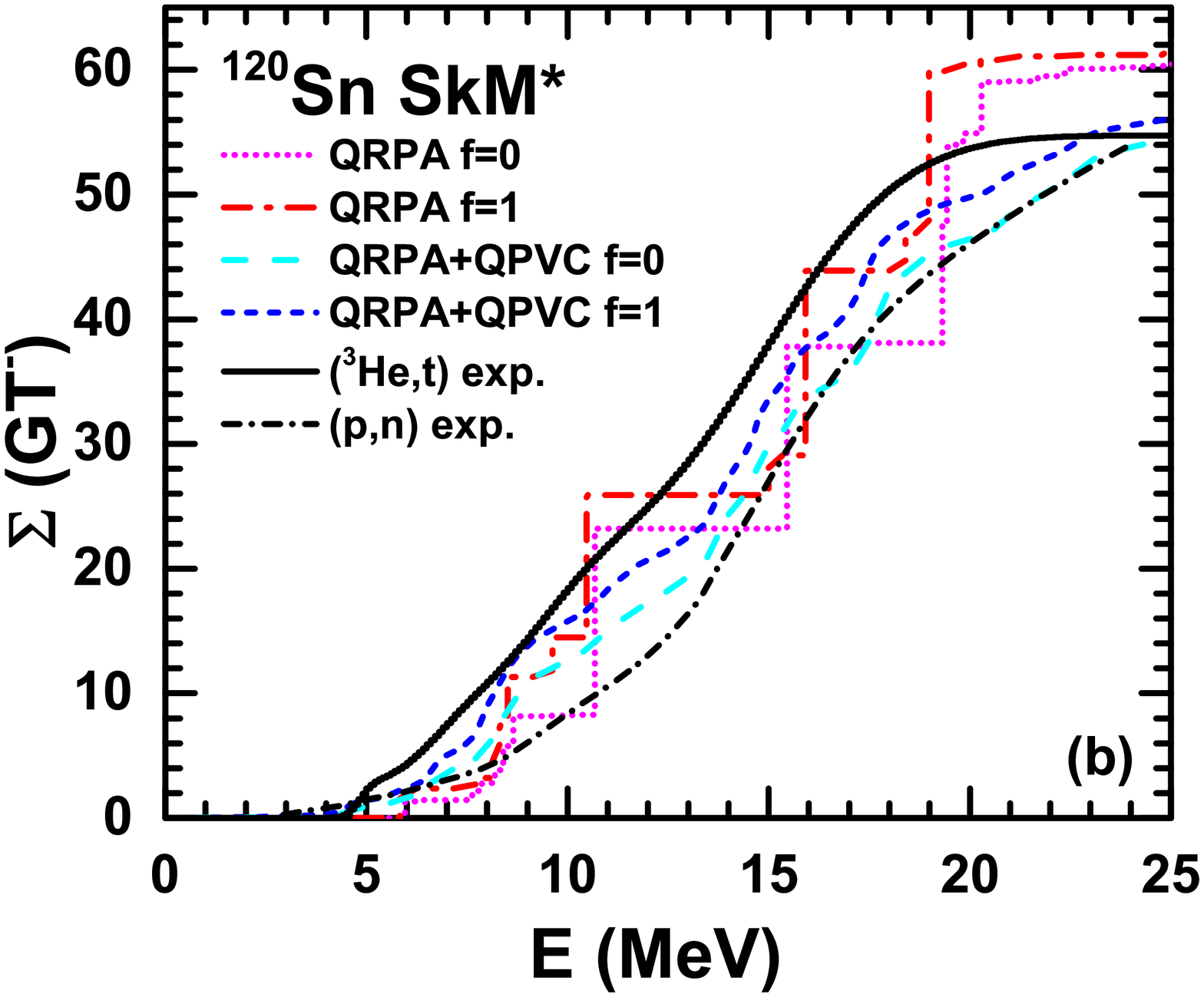}
\caption{
(Color online) The cumulative sum of Gamow-Teller strength for $^{120}$Sn, calculated by QRPA and QRPA+QPVC model, with and without isoscalar pairing, using the Skyrme interaction SkM*. The experimental results from ($^{3}$He, t) \cite{Pham1995} and (p,n) reactions \cite{Sasano2009} are shown for comparison. In panel (a), the B(GT) strength from the ($^{3}$He, t) experiment is obtained by multiplying the cross section by a factor of 1.6 so that the main GTR strength exhausts $65\%$ of Ikeda sum rule, while the B(GT) strength for (p,n) experiment is obtained by dividing the cross section by the unit cross section. In panel (b), the  two experimental cumulative GT sums are normalized to the same value as the theoretical one of QRPA+QPVC with $f=0$ at $E=25$ MeV.
}
\label{fig9}
\end{figure}

The cumulative sums of the four  strength distributions calculated by QRPA and QRPA+QPVC model with and without isoscalar pairing
are plotted in Fig. \ref{fig9}. We include for comparison  also the experimental results from ($^{3}$He, t) and (p,n) reactions.
Because the associated strength function was not given,   in panel (a) we show the ($^{3}$He, t)  cross section scaled by a factor of 1.6,
so that the main GTR strength exhausts $65$\% of the Ikeda sum rule, as reported in \cite{Pham1995}. In the work in which the (p,n) experiment has
been reported \cite{Sasano2009}, besides the  cross section $\sigma (0^o)$, the unit cross section $\hat{\sigma}$ = $2.78\pm 0.16$ mb/sr was also determined. We can obtain an approximate value for the B(GT) strength,
using the relation $\sigma (0^o)= \hat{\sigma} F(q,\omega) B(GT)$, and assuming the factor $F(q,\omega)$,  that gives the dependence on momentum and energy transfer of cross section, to be constant and equal to 1.

The resulting cumulative B(GT) is shown in panel Fig. \ref{fig9} (a). The results of these two experiments are quite different in the low-energy region and also  in the total strength up to $E=25$ MeV. In Ref. \cite{Pham1995}, it is stated that only 20\% of the observed ($^3$He, t) charge-exchange transition strength is due to $\Delta L=0$ spin-flip mediated by the central interaction $V_{\sigma\tau}$, while $\sim 80$\% is due to $\Delta L=2$ spin-flip mediated by the non-central tensor interaction $V_{T\tau}$ such as the particle-hole configurations of the type $(2d_{5/2})(1g_{7/2})^{-1}$ and $(1g_{7/2})(2d_{5/2})^{-1}$. Since the total strengths of these two experiments are not the same, we normalize the cumulative sums to the theoretical value of QRPA+QPVC with $f=0$ at $E=25$ MeV, and plot them in Fig. \ref{fig9} (b). At the QRPA level, the low-energy strength is increased going from $f=0$ to $f=1$, while the total strength at $E=25$ MeV is almost the same, and close to $3(N-Z)$. The development of the spreading width substantially improve the comparison with experiment  when going from the QRPA to QRPA+QPVC.
Going from $f=0$ to $f=1$, the empirical  low-energy strength is increased, so the $f=1$ result is more close to the ($^3$He, t) experiment. The $f=0$ result is very close to the (p,n) experiment, although it still overestimates the low-lying strength. The total strength at $E=25$ MeV is about the same with $f=0$ and $f=1$,
and is quenched by about 10\% with respect to the QRPA results.

\begin{figure}[htb]
\includegraphics[scale=0.3]{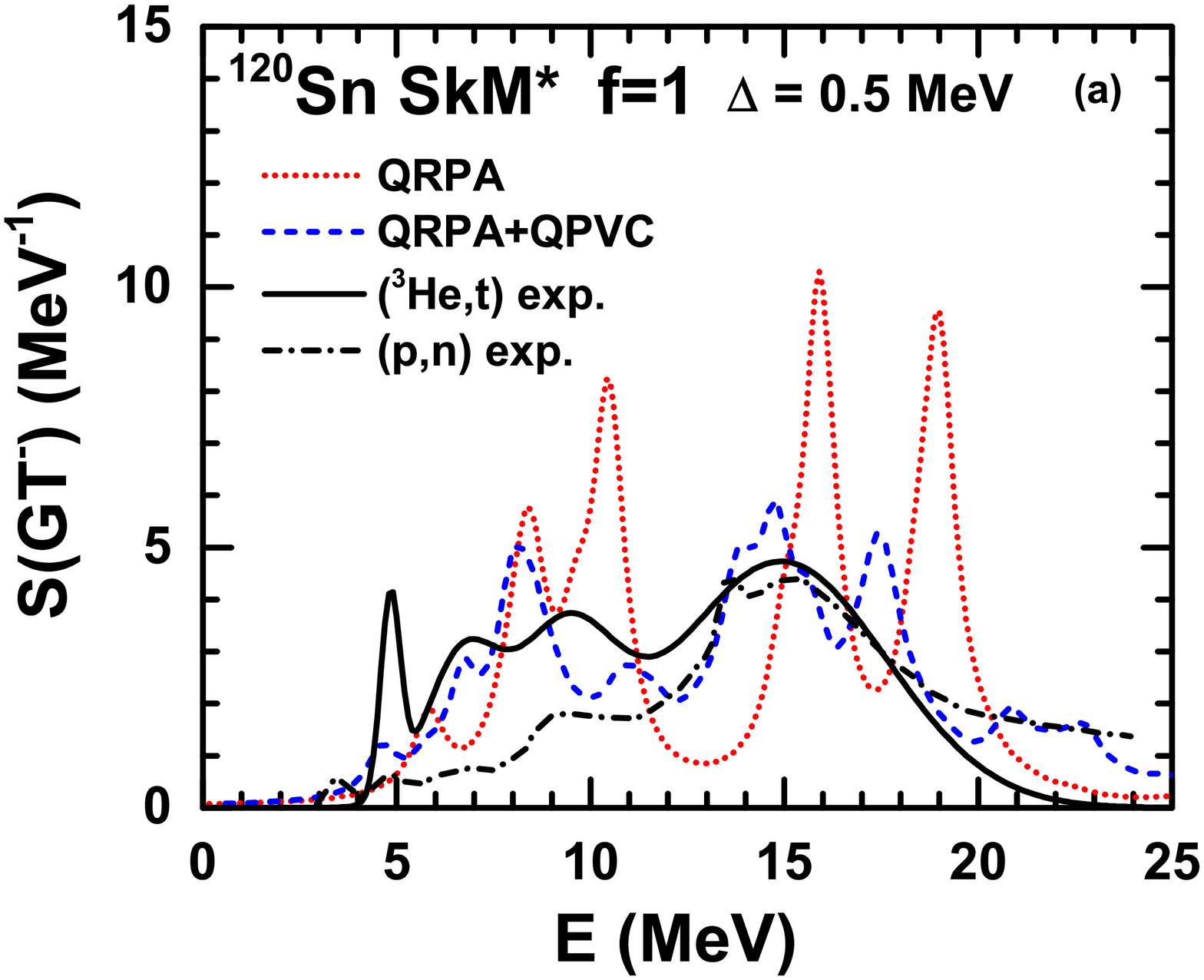}
\includegraphics[scale=0.3]{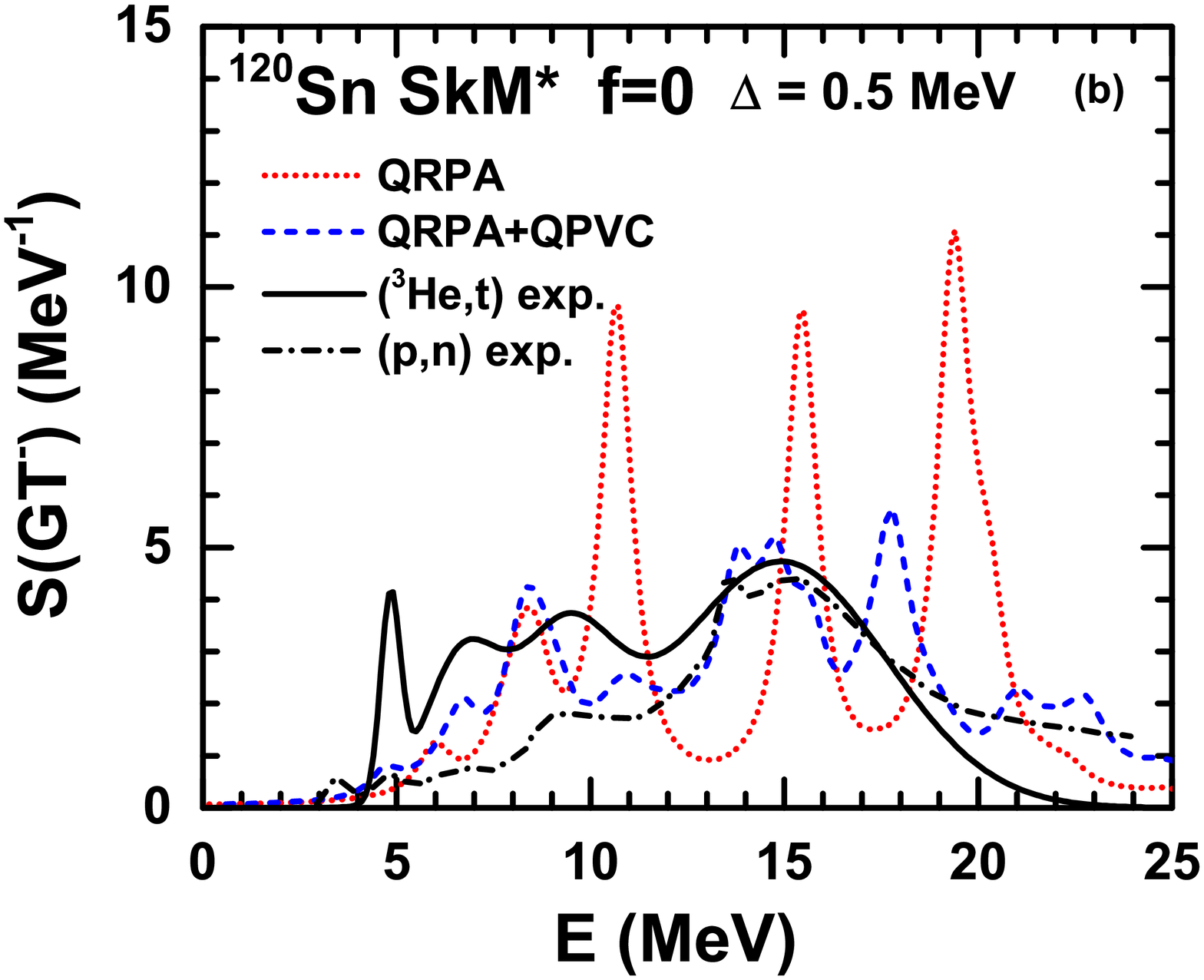}
\caption{
(Color online) The Gamow-Teller strength distributions for $^{120}$Sn calculated by QRPA and QRPA+QPVC models, with [panel (a)] and without [panel (b)] isoscalar pairing, using the Skyrme interaction SkM*.
The smearing parameter $\Delta = 0.5$ MeV is used instead of $\Delta = 0.2$ MeV used for Fig. \ref{fig8}. The experimental results from ($^{3}$He, t) and (p,n) reactions are shown for comparison. The cross section from ($^{3}$He, t) experiment is scaled by a factor of 1.6 so that the main GTR strength exhausts $65\%$ of Ikeda sum rule \cite{Pham1995}. The cross section from (p,n) reaction is normalized by the unit cross section \cite{Sasano2009} (cf. the main text).
}
\label{fig10}
\end{figure}

The four theoretical strength functions are compared with experiment in Fig. \ref{fig10}. We use a smearing parameter $\Delta=0.5$ MeV in the QRPA and QRPA+QPVC calculation,
instead of the value $\Delta =0.2 $ previously used in Fig. \ref{fig8}. This value corresponds to the energy resolution of the (p,n) experiment.
As in Fig. \ref{fig9},
the ($^{3}$He, t) experimental low-energy strength distribution is well reproduced by including isoscalar pairing, while the (p,n) data are better reproduced without it.
The spreading width and lineshape of the giant resonance region are very well reproduced by the inclusion of QPVC effect.

\section{Summary and Perspectives}
\label{summary}

The self-consistent QRPA+QPVC model based on Skyrme density functionals has been developed
for the first time
and applied to the calculation of the GT strength distribution of the superfluid nucleus $^{120}$Sn.
This model is an extension of the previously developed RPA+PVC model for magic nuclei, yet
with specific features that have been discussed in detail in this work, starting from the inclusion of isoscalar
pairing. Moreover,
the subtraction method has been adopted and its impact on the results
has been elucidated.
We have mainly discussed the results obtained by using the SkM* force, which gives the
best description among the three Skyrme forces we have considered, consistently with previous results in
non-superfluid nuclei \cite{Niu2012,Niu2014,Niu2015}.

Specifically,
the inclusion of QPVC on top of simple QRPA produces a conspicuous spreading width and
is quite relevant to reproduce well the experimental line shape of the strength distribution.
As an overall effect,
several peaks that are found in the QRPA model to lie in the giant resonance region are
merged into one big resonance peak with four subpeaks in our calculation. The microscopic structure, as well as the origin of
the widths of these four subpeaks are analyzed in detail in our paper. Eventually,
the cumulative GT strength distribution has been compared with the experimental data from ($^3$He, t) and (p,n) experiments.
Our QRPA+QPVC result is closer to the ($^3$He, t) data when isoscalar pairing is included,
while it reproduces very well the (p,n) data when this is neglected, with a slight overestimate of the low-lying strength.

The inclusion of pairing correlations paves the way to many possible applications of our model
to charge-exchange transitions
in the case of nuclei far from stability line.
In fact, the HFB plus QRPA is the appropriate tool for these neutron-rich, or neutron-deficient, nuclei, especially for
weakly bound nuclei.  Charge-exchange reactions or $\beta$-decay are valid spectroscopic tools for these nuclei, but mean-field or
DFT calculations cannot describe the damping width due to the lack of coupling with more complicated configurations,  and they also tend to overestimate the $\beta$-decay half-lives when applied to such kind of processes in exotic nuclei.
Benchmarking
PVC calculations in these cases is a new reasearch line which is still in its infancy. Improving the theoretical predictive power
of such calculations is not only beneficial for our progress in understanding nuclear structure, but also weak-interaction processes are  of essential
interest for particle physics or astrophysics. Accordingly, we envisage
the study of weak-interaction processes of astrophysical interest in our future research of QRPA+QPVC model.

{\center{\bf ACKNOWLEDGMENTS}}

This work was partly supported by the National Natural Science Foundation of China
under Grants No. 11305161 and by JSPS KAKENHI Grant Numbers JP16K05367.
Funding from the European Unions Horizon 2020 research and innovation
programme under grant agreement No. 654002 is also
acknowledged.


\section{Appendix}

 \subsection{The calculation of $\langle ab | V | N \rangle$}
 \label{appendixA}

We wish to calculate
 \begin{equation}
 \langle ab | V | N \rangle =  \langle 0 | \alpha_{b} \alpha_{a} V \alpha_{a''} ^\dagger \alpha_{b''} ^\dagger \Gamma_{nL}^\dagger | 0\rangle,
\end{equation}
with
\begin{equation}
  \Gamma_{nL}^\dagger = \frac{1}{\sqrt{1+\delta_{cd}}} \sum_{c \geq d} X_{cd} \alpha_{c} ^\dagger \alpha_{d} ^\dagger - Y_{cd} \alpha_d \alpha_c.
 \end{equation}
Since $|0\rangle$ is the vacuum for HFB quasi-particle states, we transform the BCS quasi-particle states $|a\rangle$, associated
with the operators $\alpha^\dagger_{{a}}$, to HFB quasi-particle states $|\tilde{a}\rangle$, associated with the operators $\beta^\dagger_{\tilde{a}}$, based on the following unitary transformation \cite{RingBook},
\begin{eqnarray}
 \alpha_{{a}}  &=& \sum_{\tilde{a}} C_{a \tilde{a}} \beta_{\tilde{a}} , \\
 \alpha^\dagger_{{a}} & =& \sum_{\tilde{a}} C^*_{ a \tilde{a}  } \beta_{\tilde{a}}^\dagger .
\end{eqnarray}
We will have
\begin{equation}
 \langle ab | V | N \rangle = \sum_{\tilde{a}\tilde{b}\tilde{a}''\tilde{b}''} C_{a \tilde{a}} C_{b\tilde{b}} C^*_{a''\tilde{a}''} C^*_{b''\tilde{b}''}\langle 0 | \beta_{\tilde{b}} \beta_{\tilde{a}} V \beta_{\tilde{a}''} ^\dagger \beta_{\tilde{b}''} ^\dagger \Gamma_{nL}^\dagger | 0\rangle,
\end{equation}
and
\begin{equation}
 \Gamma_{nL}^\dagger = \frac{1}{\sqrt{1+\delta_{cd}}} \sum_{c \geq d} \sum_{\tilde{c}\tilde{d}}  C^*_{c\tilde{c} } C^*_{d\tilde{d}}X_{cd} \beta_{\tilde{c}} ^\dagger \beta_{\tilde{d}} ^\dagger - C_{c\tilde{c}}C_{d\tilde{d}} Y_{cd}  \beta_{\tilde{d}} \beta_{\tilde{c}}.
\end{equation}
 In the case of QRPA phonons, we make the following approximation,
 \begin{equation}
\langle 0 | \beta_{\tilde{b}} \beta_{\tilde{a}} V \beta_{\tilde{a}''} ^\dagger \beta_{\tilde{b}''} ^\dagger \Gamma_{nL}^\dagger | 0\rangle  \simeq      \langle 0 | \beta_{\tilde{b}} \beta_{\tilde{a}} [V, \Gamma_{nL}^\dagger]  \beta_{\tilde{a}''} ^\dagger \beta_{\tilde{b}''} ^\dagger  | 0\rangle.
\end{equation}
 Then we obtain
 \begin{eqnarray}
  \langle ab | V | N \rangle &=& \frac{1}{\sqrt{1+\delta_{cd}}} \sum_{\tilde{a}\tilde{b}\tilde{a}''\tilde{b}''} C_{a \tilde{a}} C_{b\tilde{b}} C^*_{a''\tilde{a}''} C^*_{b''\tilde{b}''}\sum_{c\geq d}\sum_{\tilde{c}\tilde{d}}  \nonumber\\
  &&  [C^*_{c\tilde{c} } C^*_{d\tilde{d}}\langle 0 | \beta_{\tilde{b}} \beta_{\tilde{a}} V \beta_{\tilde{a}''} ^\dagger \beta_{\tilde{b}''} ^\dagger \beta_{\tilde{c}} ^\dagger \beta_{\tilde{d}} ^\dagger | 0\rangle  X_{cd} + C_{c\tilde{c}}C_{d\tilde{d}}\langle 0 | \beta_{\tilde{b}} \beta_{\tilde{a}}   \beta_{\tilde{d}}\beta_{\tilde{c}}   V   \beta_{\tilde{a}''} ^\dagger \beta_{\tilde{b}''} ^\dagger  | 0\rangle  Y_{cd}].
\end{eqnarray}

 Here $V$ is the two-body interaction for the coupling vertex. It has the general form in the single-particle basis,
  \begin{equation}
  V = \frac{1}{4} \sum_{1234} {V}_{1234} c^\dagger_{1} c^\dagger_{2} c_{4} c_{3}  ,
  \end{equation}
  and can be written in the HFB quasi-particle basis,
  \begin{equation}
   V = \sum_{\tilde{a}\tilde{b}\tilde{c}\tilde{d}} ( H_{\tilde{a}\tilde{b}\tilde{c}\tilde{d}}^{40} \beta_{\tilde{a}}^\dagger \beta_{\tilde{b}}^\dagger \beta_{\tilde{c}}^\dagger \beta_{\tilde{d}}^\dagger  + h.c.)
     +  \sum_{\tilde{a}\tilde{b}\tilde{c}\tilde{d}} ( H_{\tilde{a}\tilde{b}\tilde{c}\tilde{d}}^{31} \beta_{\tilde{a}}^\dagger \beta_{\tilde{b}}^\dagger \beta_{\tilde{c}}^\dagger \beta_{\tilde{d}}   + h.c.)
     +  \frac{1}{4} \sum_{\tilde{a}\tilde{b}\tilde{c}\tilde{d}} H_{\tilde{a}\tilde{b}\tilde{c}\tilde{d}}^{22} \beta_{\tilde{a}}^\dagger \beta_{\tilde{b}}^\dagger \beta_{\tilde{d}} \beta_{\tilde{c}} .
 \end{equation}

 Using the Wick theorem, only the $H^{31}$ or $H^{13}$ terms exist in $\langle ab | V | N \rangle$, and we get

\begin{eqnarray}
\label{eqVabN}
 \langle ab | V | N \rangle &=& \frac{1}{\sqrt{1+\delta_{cd}}} \sum_{\tilde{a}\tilde{b}\tilde{a}''\tilde{b}''}  C_{a \tilde{a}} C_{b\tilde{b}} C^*_{a''\tilde{a}''} C^*_{b''\tilde{b}''} \sum_{c\geq d} \sum_{\tilde{c}\tilde{d}} \nonumber\\
  &&C^*_{c\tilde{c} } C^*_{d\tilde{d}} [ \delta_{\tilde{b}\tilde{b}''}  ( H^{13}_{\tilde{a}''\tilde{c}\tilde{d}\tilde{a}} - H^{13}_{\tilde{a}''\tilde{d}\tilde{c}\tilde{a}} + H^{13}_{\tilde{c}\tilde{d}\tilde{a}''\tilde{a}} - H^{13}_{\tilde{c}\tilde{a}''\tilde{d}\tilde{a}} + H^{13}_{\tilde{d}\tilde{a}''\tilde{c}\tilde{a}} - H^{13}_{\tilde{d}\tilde{c}\tilde{a}''\tilde{a}}) X_{cd} \nonumber\\
  & & + \delta_{\tilde{a}\tilde{a}''}  ( H^{13}_{\tilde{b}''\tilde{c}\tilde{d}\tilde{b}} - H^{13}_{\tilde{b}''\tilde{d}\tilde{c}\tilde{b}} + H^{13}_{\tilde{c}\tilde{d}\tilde{b}''\tilde{b}} - H^{13}_{\tilde{c}\tilde{b}''\tilde{d}\tilde{b}} + H^{13}_{\tilde{d}\tilde{b}''\tilde{c}\tilde{b}} - H^{13}_{\tilde{d}\tilde{c}\tilde{b}''\tilde{b}}) X_{cd} ] \nonumber\\
  && +C_{c\tilde{c}}C_{d\tilde{d}} [\delta_{\tilde{b}\tilde{b}''} ( H^{31}_{\tilde{a}\tilde{c}\tilde{d}\tilde{a}''} - H^{31}_{\tilde{a}\tilde{d}\tilde{c}\tilde{a}''} + H^{31}_{\tilde{c}\tilde{d}\tilde{a}\tilde{a}''} - H^{31}_{\tilde{c}\tilde{a}\tilde{d}\tilde{a}''} + H^{31}_{\tilde{d}\tilde{a}\tilde{c}\tilde{a}''} - H^{31}_{\tilde{d}\tilde{c}\tilde{a}\tilde{a}''})  Y_{cd}  \nonumber\\
  & & + \delta_{\tilde{a}\tilde{a}''}   ( H^{31}_{\tilde{b}\tilde{c}\tilde{d}\tilde{b}''} - H^{31}_{\tilde{b}\tilde{d}\tilde{c}\tilde{b}''} + H^{31}_{\tilde{c}\tilde{d}\tilde{b}\tilde{b}''} - H^{31}_{\tilde{c}\tilde{b}\tilde{d}\tilde{b}''} + H^{31}_{\tilde{d}\tilde{b}\tilde{c}\tilde{b}''} - H^{31}_{\tilde{d}\tilde{c}\tilde{b}\tilde{b}''})   Y_{cd} ]\\
  &=& \frac{1}{\sqrt{1+\delta_{cd}}} \sum_{\tilde{a}\tilde{b}\tilde{a}''\tilde{b}''}  C_{a \tilde{a}} C_{b\tilde{b}} C^*_{a''\tilde{a}''} C^*_{b''\tilde{b}''} \sum_{c\geq d} \sum_{\tilde{c}\tilde{d}} \nonumber\\
  &&C^*_{c\tilde{c} } C^*_{d\tilde{d}} [ \delta_{\tilde{b}\tilde{b}''} 2 ( H^{13}_{\tilde{a}''\tilde{c}\tilde{d}\tilde{a}} + H^{13}_{\tilde{c}\tilde{d}\tilde{a}''\tilde{a}}  + H^{13}_{\tilde{d}\tilde{a}''\tilde{c}\tilde{a}} )X_{cd} \nonumber\\
  & & + \delta_{\tilde{a}\tilde{a}''} 2 ( H^{13}_{\tilde{b}''\tilde{c}\tilde{d}\tilde{b}} + H^{13}_{\tilde{c}\tilde{d}\tilde{b}''\tilde{b}} + H^{13}_{\tilde{d}\tilde{b}''\tilde{c}\tilde{b}} ) X_{cd} ] \nonumber\\
  && +C_{c\tilde{c}}C_{d\tilde{d}} [\delta_{\tilde{b}\tilde{b}''} 2( H^{31}_{\tilde{a}\tilde{c}\tilde{d}\tilde{a}''}  + H^{31}_{\tilde{c}\tilde{d}\tilde{a}\tilde{a}''} + H^{31}_{\tilde{d}\tilde{a}\tilde{c}\tilde{a}''} )  Y_{cd}  \nonumber\\
  & & + \delta_{\tilde{a}\tilde{a}''}   ( H^{31}_{\tilde{b}\tilde{c}\tilde{d}\tilde{b}''}  + H^{31}_{\tilde{c}\tilde{d}\tilde{b}\tilde{b}''} + H^{31}_{\tilde{d}\tilde{b}\tilde{c}\tilde{b}''} )   Y_{cd} ].
\end{eqnarray}

With the $C$ matrix, $H^{31}_{\tilde{a}\tilde{b}\tilde{c}\tilde{d}}$ on HFB quasi-particle states can be transformed to $H^{31}_{abcd}$ on BCS quasi-particle states, so that

\begin{eqnarray}
\label{eqVabN2}
 \langle ab | V | N \rangle &=& \frac{1}{\sqrt{1+\delta_{cd}}} \sum_{c\geq d} [ \delta_{bb''} 2 ( H^{13}_{a''cda}  + H^{13}_{cda''a} + H^{13}_{da''ca} ) X_{cd} \nonumber\\
  & & + \delta_{aa''} 2 ( H^{13}_{b''cdb}  + H^{13}_{cdb''b}  + H^{13}_{db''cb} ) X_{cd} ] \nonumber\\
  && + [\delta_{bb''} 2( H^{31}_{acda''}  + H^{31}_{cdaa''}  + H^{31}_{daca''} )  Y_{cd}  \nonumber\\
  & & + \delta_{aa''}  2 ( H^{31}_{bcdb''}  + H^{31}_{cdbb''}  + H^{31}_{dbcb''} )   Y_{cd} ].
\end{eqnarray}

From Ref. \cite{RingBook} we know that
\begin{equation}
 H^{31}_{\tilde{a}\tilde{b}\tilde{c}\tilde{d}} = \frac{1}{2} \sum_{1234} {V}_{1234} [U^*_{1\tilde{a}} V^*_{4\tilde{b}} V^*_{3\tilde{c}} V_{2\tilde{d}} + V_{3\tilde{a}}^* U_{2\tilde{b}}^* U_{1\tilde{c}}^* U_{4\tilde{d}}],
\end{equation}
where $\tilde{a},\tilde{b},\tilde{c},\tilde{d}$ denote the states of the quasi-particle basis, and $1,2,3,4$ denote the states of the single-particle basis. After transformation with $C$, $H^{31}_{\tilde{a}\tilde{b}\tilde{c}\tilde{d}}$ becomes $H^{31}_{abcd}$ on the BCS quasi-particle basis with the form,

\begin{equation}
 H^{31}_{abcd} = \frac{1}{2} \sum_{\tilde{1}\tilde{2}\tilde{3}\tilde{4}} {V}_{\tilde{1}\tilde{2}\tilde{3}\tilde{4}} [\bar{U}^*_{\tilde{1}{a}} \bar{V}^*_{\tilde{4}{b}} \bar{V}^*_{\tilde{3}{c}} \bar{V}_{\tilde{2}{d}} + \bar{V}_{\tilde{3}{a}}^* \bar{U}_{\tilde{2}{b}}^* \bar{U}_{\tilde{1}{c}}^* \bar{U}_{\tilde{4}{d}}],
\end{equation}
where $\tilde{1},\tilde{2},\tilde{3},\tilde{4}$ denote the canonical basis.  The $\bar{U}$ and $\bar{V}$ matrices connect the canonical basis and BCS quasi-particle basis, and their definition is found  in Ref. \cite{RingBook}. The $\bar{U}$ and $\bar{V}$ matrices can be further simplified as
\begin{equation}
 \bar{U}_{\tilde{1}a} = u_{\tilde{1}} \delta_{\tilde{1}a}, \quad \bar{V}_{\tilde{1}a} = - v_{\tilde{1}}\delta_{\tilde{1}\bar{a}},
\end{equation}
where  $u,v$ denote the occupation amplitudes in the canonical basis.
Then

\begin{eqnarray}
 H^{31} &=& \sum_{abcd} H^{31}_{abcd} \alpha_a^\dagger \alpha_b^\dagger \alpha_c^\dagger \alpha_d \nonumber\\
 &=& \sum_{abcd} \frac{1}{2} \sum_{\tilde{1}\tilde{2}\tilde{3}\tilde{4}} {V}_{\tilde{1}\tilde{2}\tilde{3}\tilde{4}} [\bar{U}^*_{\tilde{1}{a}} \bar{V}^*_{\tilde{4}{b}} \bar{V}^*_{\tilde{3}{c}} \bar{V}_{\tilde{2}{d}} + \bar{V}_{\tilde{3}{a}}^* \bar{U}_{\tilde{2}{b}}^* \bar{U}_{\tilde{1}{c}}^* \bar{U}_{\tilde{4}{d}}]  \alpha_a^\dagger \alpha_b^\dagger \alpha_c^\dagger \alpha_d \nonumber\\
 &=& - \sum_{abcd} \frac{1}{2} \sum_{\tilde{1}\tilde{2}\tilde{3}\tilde{4}} {V}_{\tilde{1}\tilde{2}\tilde{3}\tilde{4}} (u_{\tilde{1}} v_{\tilde{2}} v_{\tilde{3}} v_{\tilde{4}} \delta_{\tilde{1}a} \delta_{\tilde{2}\bar{d}} \delta_{\tilde{3}\bar{c}} \delta_{\tilde{4}\bar{b}} \alpha^\dagger_{\tilde{1}} \alpha^\dagger_{\bar{\tilde{4}}} \alpha^\dagger_{\bar{\tilde{3}}}\alpha_{\bar{\tilde{2}}}+  u_{\tilde{1}}u_{\tilde{2}}v_{\tilde{3}}u_4 \delta_{\tilde{1}c}\delta_{\tilde{2}b} \delta_{\tilde{3}\bar{a}} \delta_{\tilde{4}d} \alpha^\dagger_{\bar{\tilde{3}}}\alpha^\dagger_{\tilde{2}}\alpha^\dagger_{\tilde{1}}\alpha_{\tilde{4}}) \nonumber\\
 &=& \sum_{abcd}\frac{1}{2} \sum_{\tilde{1}\tilde{2}\tilde{3}\tilde{4}} {V}_{\tilde{4}\bar{\tilde{3}}\tilde{1}\tilde{2}} [ u_{\tilde{1}}u_{\tilde{2}} v_{ \tilde{3}}  u_{\tilde{4}} - v_{ \tilde{1}} v_{ \tilde{2}}  u_{\tilde{3}}  v_{ \tilde{4}}]  \alpha^\dagger_{\tilde{1}} \alpha^\dagger_{\tilde{2}} \alpha^\dagger _{\tilde{3}} \alpha_{\tilde{4}} \delta_{\tilde{1}a} \delta_{\tilde{2}b}\delta_{\tilde{3}c}\delta_{\tilde{4}d} \\
 &=& \frac{1}{2} \sum_{\tilde{1}\tilde{2}\tilde{3}\tilde{4}} {V}_{\tilde{4}\bar{\tilde{3}}\tilde{1}\tilde{2}} [ u_{\tilde{1}}u_{\tilde{2}} v_{ \tilde{3}}  u_{\tilde{4}} - v_{ \tilde{1}} v_{ \tilde{2}}  u_{\tilde{3}}  v_{ \tilde{4}}]  \alpha^\dagger_{\tilde{1}} \alpha^\dagger_{\tilde{2}} \alpha^\dagger _{\tilde{3}} \alpha_{\tilde{4}} \\
 &=&\frac{1}{2} \sum_{abcd}  {V}_{d\bar{c}ab} [ u_{a}u_{b} v_{ c}  u_{d} - v_{ a} v_{ b}  u_{c}  v_{ d}]  \alpha^\dagger_{a} \alpha^\dagger_{b} \alpha^\dagger _{c} \alpha_{d}.
\end{eqnarray}
Finally we can write $H^{31}_{abcd}$ in the canonical basis,
 \begin{equation}
 \label{eqH31}
 H^{31}_{abcd}= \frac{1}{2}  {V}_{d\bar{c}ab} [
          u_{a}u_{b} v_{ c}  u_{d} -  v_{ a} v_{ b}  u_{c}  v_{ d}].
 \end{equation}
 Similarly,

   \begin{equation}
   \label{eqH13}
  H^{13}_{abcd} =   \frac{1}{2}    {V}^*_{d\bar{c}ab}  [  u_{ a}  u_{b}  v_{ c}   u_{d} -  v _{a} v _{ b }  u _{c}  v_{d} ] .
  \end{equation}

   Combining the above Eq. (\ref{eqH31}) and Eq. (\ref{eqH13}) with Eq. (\ref{eqVabN2}), one arrives at Eq. (\ref{eqWdown2}) and (\ref{eqWdown3}) in Sec. \ref{formalism}.

\subsection{Angular momentum coupled form}
\label{appendixB}

In Eq. (\ref{eqVabN2}), the term $V(aa''cd)$ in front of $X$ can be expressed in angular momentum coupled form,

\begin{equation}
   V(aa''cd) =    \langle j_c m_c j_d m_d | LM \rangle \langle j_a m_a j_{a''} -m_{a''} | LM \rangle (-1)^{j_{a''}-m_{a''}} \frac{\hat{L}}{\hat{j}_a} V(cd L a'';a),
 \end{equation}
 where
 \begin{eqnarray}
  V(cd L a'';a) &=& \sum_{m_cm_dm_{a''} m_a } \langle j_c m_c j_d m_d | LM \rangle \langle j_{a''} m_{a''} LM | j_a m_a\rangle V(aa''cd) \\
                  &=& \sum_{m_cm_dm_{a''} m_a } \langle j_c m_c j_d m_d | LM \rangle \langle j_am_a j_{a''} -m_{a''} | LM \rangle (-1)^{j_{a''}-m_{a''}} \frac{\hat{j}_a}{\hat{L}}  V(aa''cd).\nonumber\\
 \end{eqnarray}
Calculating the three matrix elements in $V(aa''cd)$ with the
Clebsch-Gordan coefficients, we finally get
\begin{eqnarray}
   V(cd L a'';a)
                 &=&  \frac{\hat{j}_a}{\hat{L}} [ {V}^{L ph}_{a d a''c}  (u_{a''}u_cv_{ {d}}u_a -v_{ {a}''}v_{ {c}}u_dv_{ {a}}) \nonumber\\
                 && +    {V}^{L ph}_{aca''d}  ( u_{d}u_{a''}v_{ {c}}u_a-v_{ {d}}v_{ {a}''}u_cv_{ {a}} ) (-1)^{j_c-j_d+L} \nonumber\\
  &&  -   {V}^{L pp}_{aa''cd} (  u_{c}u_dv_{ {a}''}u_a -v_{ {c} }v_{ {d}}u_{a''}v_{ {a}})]\\
  &\equiv&  \widetilde{V}(cd L a'';a)  \frac{\hat{j}_a}{\hat{L}}.
 \end{eqnarray}
 Similarly, the term in front of $Y$ is
  \begin{equation}
   V(a''acd) =   \langle j_c m_c j_d m_d | L-M \rangle (-1)^{L-M} \langle j_a m_a j_{a''} -m_{a''} | LM \rangle (-1)^{j_{a''}-m_{a''}} \frac{\hat{L}}{\hat{j}_a} V(cd L a;a''),
 \end{equation}
 where
 \begin{eqnarray}
  V(cd L a;a'') &=& \sum_{m_cm_dm_{a''} m_a } \langle j_c m_c j_d m_d | L-M \rangle (-1)^{L-M} \langle j_{a''} m_{a''} LM | j_a m_a\rangle V(a''acd) \\
                  &=& \sum_{m_cm_dm_{a''} m_a } \langle j_c m_c j_d m_d | L-M \rangle (-1)^{L-M} \langle j_am_a j_{a''} -m_{a''} | LM \rangle (-1)^{j_{a''}-m_{a''}} \nonumber\\
                  && \frac{\hat{j}_a}{\hat{L}}  V(a''acd).
 \end{eqnarray}
 We have
  \begin{eqnarray}
   V(cd L a;a'')                &=&   [ {V}^{L ph}_{a'' d a c}  (u_{a''}u_cv_{ {d}}u_a -v_{ {a}''}v_{ {c}}u_dv_{ {a}} ) \nonumber\\
                  && +    {V}^{L ph}_{a''cad}  ( u_{d}u_{a''}v_{ {c}}u_a -v_{ {d}}v_{ {a}''}u_cv_{ {a}}) (-1)^{j_c-j_d+L} \nonumber\\
  &&  -   {V}^{L pp}_{a''acd}  ( u_{c}u_dv_{ {a}}u_{a''}-v_{ {c} }v_{ {d}}u_{a}v_{ {a}''} )](-1)^{j_a-j_{a''}+L} \frac{\hat{j}_a}{\hat{L}}\\
  & \equiv &  \widetilde{V}(cd L a;a'') (-1)^{j_a-j_{a''}+L} \frac{\hat{j}_a}{\hat{L}}.
 \end{eqnarray}

 The $X$ and $Y$ can also be written in the angular momentum coupled form,
 \begin{eqnarray}
  X_{cd} &=& \sum_{L'M'} \langle j_c m_c j_d m_d | L'M'\rangle X_{cd}^{L'M'} , \\
  Y_{cd} &=& \sum_{L'M'} \langle j_c m_c j_d m_d | L'-M'\rangle (-1)^{L'-M'} Y_{cd}^{L'M'}.
  \end{eqnarray}

  So finally $\langle a'', nL | V | a \rangle$ in angular momentum coupled form is
     \begin{eqnarray}
     \label{eqWdown1J}
  && \langle a'', nL | V | a \rangle = \langle a  | V | a'', nL \rangle = \frac{1}{\sqrt{1+\delta_{cd}}} \sum_{c\geq d} [ V(aa''cd) X_{cd} + V(a''acd) Y_{cd}  ]  \nonumber\\
  &= & \frac{1}{\sqrt{1+\delta_{cd}}}[\sum_{L'M'}  \sum_{j_c j_d  }   \langle j_a m_a j_{a''} -m_{a''} | LM \rangle (-1)^{j_{a''}-m_{a''}} \frac{\hat{L}}{\hat{j}_a} V(cd L a'';a)   \delta_{LL'} \delta_{MM'} X_{cd}^{LM}\nonumber\\
   && + \sum_{L'M'} \sum_{j_c j_d  }     (-1)^{L-M} \langle j_a m_a j_{a''} -m_{a''} | LM \rangle (-1)^{j_{a''}-m_{a''}} \frac{\hat{L}}{\hat{j}_a} V(cd L a;a'')   \delta_{LL'} \delta_{MM'}   Y_{cd}^{LM} ] \nonumber\\
  &=& \frac{1}{\sqrt{1+\delta_{cd}}}  \sum_{j_c j_d  }  \langle j_a m_a j_{a''} -m_{a''} | LM \rangle (-1)^{j_{a''}-m_{a''}} \nonumber\\
  && [\widetilde{V}(cd L a'';a)   X_{cd}^{LM} + (-1)^{j_a-j_{a''}+L} \widetilde{V}(cd L a;a'')     Y_{cd}^{LM}]  \\
  &\equiv &     \langle j_a m_a j_{a''} -m_{a''} | LM \rangle (-1)^{j_{a''}-m_{a''}}  \langle a'', nL || V || a \rangle \\
  &=&     \langle j_a m_a j_{a''} -m_{a''} | LM \rangle (-1)^{j_{a''}-m_{a''}}\langle a  || V || a'', nL \rangle.
 \end{eqnarray}

 With the above expressions, we can obtain the angular momentum coupled form of $\langle ab | V | N \rangle$, and hence the $W^{\downarrow}_{ab,a'b'}$. Through the following relation,
 \begin{equation}
 W^{\downarrow J}_{ aba'b'}  = \sum_{m_am_bm_{a'} m_{b'}} \langle j_am_a j_{b} m_{b} | JM_J  \rangle \langle j_{a'} m_{a'}j_{b'} m_{b'} | JM_J \rangle  W^{\downarrow}_{aba'b'},
 \end{equation}
the angular momentum coupled $W^{\downarrow J}_{ab,a'b'}$ in Eq. (\ref{Wdown1}) will be obtained.
%

\clearpage


\end{document}